%% file: ms.tex
\documentclass[preprint,12pt]{elsarticle}

\usepackage{amssymb}

\usepackage{hyperref}
\usepackage{xcolor}
\newcommand{\mv}{MultiVeStA}

\newcommand{\rev}[1]{#1}
\newcommand{\revsecond}[1]{\textcolor{black}{#1}}
\newcommand{\rtwo}[1]{\revsecond{#1}}
\newcommand{\codefont}[1]{{\small\texttt{#1}}}
\newcommand{\cf}[1]{\codefont{#1}}
\usepackage{soul}
\usepackage[T1]{fontenc}
\usepackage{todonotes}
\usepackage{soul}
\usepackage{graphicx}
\usepackage{listings}
\usepackage{wrapfig}
\usepackage{caption}
\usepackage{subcaption}
\usepackage{amsmath}
\usepackage{xfrac}
\usepackage{xcolor}
\usepackage{booktabs}
\usepackage{adjustbox}

\journal{Journal of Systems and Software}

\definecolor{keyworkColor}{rgb}{0.5,0,0.35}
\definecolor{commentColor}{RGB}{63,127,95}

\begin{document}

\begin{frontmatter}



\title{White-box validation of \rev{quantitative 
product lines} by statistical model checking and process mining}


\address[sssa]{Institute of Economics and L'EMbeDS, Sant'Anna School of Advanced Studies, Pisa, Italy.}
\address[dtu]{DTU Technical University of Denmark, Lyngby, Denmark.}
\address[pennstate]{Dept. of Statistics and Huck Institutes of the Life Sciences, Penn State University, USA}

\author[sssa]{Roberto Casaluce}
\ead{roberto.casaluce@santannapisa.it}
\author[dtu]{Andrea Burattin}
\ead{andbur@dtu.dk}
\author[sssa,pennstate]{Francesca Chiaromonte}
\ead{francesca.chiaromonte@santannapisa.it}
\author[dtu]{Alberto Lluch Lafuente}
\ead{albl@dtu.dk}
\author[sssa,dtu]{Andrea Vandin\corref{cor1}}
\ead{andrea.vandin@santannapisa.it}
\cortext[cor1]{Corresponding author}

\begin{abstract}
We propose a novel methodology to validate software product line (PL) models by integrating Statistical Model Checking (SMC) with Process Mining (PM). 
We consider the feature-oriented language QFLan from the PL engineering domain. QFLan allows to model \rev{PL equipped with rich cross-tree and quantitative constraints, as well as aspects of dynamic PLs
such as the staged configurations}. 
\rtwo{This richness allows us to easily obtain models with infinite state-space, calling for simulation-based analysis techniques, like SMC. For example, we use a running example with infinite state space.}
SMC is a family of analysis techniques 
based on the generation of samples of the dynamics of a system. SMC aims at estimating properties of a system like the probability of a given event (e.g., installing a feature), or the expected value of quantities in it (e.g., the \rev{average} price of products from the studied family).  Instead, PM is a family of data-driven techniques that uses logs collected on the execution of an information system   to identify and reason about its underlying execution process. This often regards identifying and reasoning about process patterns, bottlenecks, and possibilities for improvement. \rtwo{In this paper, to the best of our knowledge, we propose, for the first time, the application of Process Mining (PM) techniques to the byproducts of Statistical Model Checking (SMC) simulations. This aims to enhance the utility of SMC analyses.}

Typically, if  SMC gives unexpected results, the modeler has to discover whether these come from actual characteristics of the system, or from bugs in the model. This is  done in a \emph{black-box} manner, only based on the obtained numerical values. We improve on this by using PM to get a white-box perspective on the dynamics of the system observed by SMC. Roughly speaking, we feed the samples generated by SMC to PM tools, obtaining 
a compact graphical representation of the observed dynamics. This \emph{mined PM model} is then transformed into a \emph{mined QFLan model}, making it  accessible to PL engineers. 
%
Using two well-known PL models, we show that our methodology is effective (helps in pinpointing issues in models, and in suggesting fixes), and that it  scales to complex models. We also show that it is general, by applying it to the security domain. 
%
\end{abstract}


\begin{highlights}
\item Effective, scalable, and multi-domain methodology for the analysis of product line models
\item \rev{Integration of statistical model checking and process mining applied to product line and security domains}
\item From black-box model validation to white-box novel validation and enhancement 
\item Mining of product line models
\end{highlights}


\end{frontmatter}


\input{intro}

\input{background}

\input{method}

\input{validation}

\input{relatedwork}
\input{discussion}






\bibliographystyle{elsarticle-num} 
\bibliography{bibitems}

\section{Vitae}

\textbf{Roberto Casaluce} is a PhD student in Artificial Intelligence for Society at the University of Pisa and the Sant'Anna School of Advanced Studies, Pisa. He completed an MSc in Politics and an MSc in Data Science at Birkbeck, University of London. For his latter master’s thesis he worked on a project within the ETP Group of ETH Zurich using NLP methods to classify Clean Energy Storage Technologies patents. After submitting his master’s thesis, he worked as a research assistant on the same project for six months. His research interests include statistical model checking techniques, process mining, and NLP.

\textbf{Andrea Burattin} is Associate Professor at the Technical University of Denmark since April 2019. Previously, he worked as Assistant Professor at the same university, and as postdoctoral researcher at the University Innsbruck (Austria) and at the University of Padua (Italy). In 2013 he obtained his PhD degree from a joint PhD School between the University of Bologna and Padua (Italy). His PhD thesis received the Best Process Mining Dissertation Award from the IEEE Task Force on Process Mining. He is member of the steering committee of the IEEE Task Force on Process Mining.

\textbf{Francesca Chiaromonte} is a Professor of Statistics at the Sant’Anna School of Advanced Studies, where she is the Scientific Coordinator of EMbeDS – a department of excellence fostering data- and computation-intensive research in the social sciences. She is a member of the board of the National Italian PhD in AI and Society, and she holds the Dorothy Foehr Huck and J. Lloyd Huck Chair in Statistics for the Life Sciences at the Pe nnsylvania State University.  Francesca is a fellow of the American Statistical Association since 2016, and a fellow of the Institute for Mathematical Statistics since 2022.

\textbf{Alberto Lluch Lafuente} is Full Professor at the Technical University of Denmark. He is currently head of the section for Software Systems Engineering. Previously, he was Associate Professor at the Technical University of Denmark, Assistant Professor at IMT School for Advanced Studies Lucca and Postdoctoral Researcher at the University of Pisa. He obtained his PhD degree from the Albert-Ludwigs Universit\"at in Freiburg in 2003. 
 
\textbf{Andrea Vandin} is Associate Professor in Computer Science at Sant'Anna School of Advanced Studies, Pisa, since December 2022, as well as Adjunct Associate Professor at DTU Technical University of Denmark. Previously, he worked as Associate Professor at DTU, and Assistant Professor at IMT School for Advanced Studies Lucca, and the University of Southampton UK. In 2013, he obtained his PhD degree in Computer Science and Engineering at IMT. He is member of the board of the National Italian PhD in AI and Society.\\

\input{appendix}

%
%
%
%

\end{document}

%% file: intro.tex
\section{Introduction}
\label{sec:intro}


Software product lines (SPL), and feature models in general, as well as Product Line Engineering (PLE), play a very important role in modern society, where customization capabilities are expected even for commodity products. Very often, these products are equipped with software that is expected to follow the customization of the product itself. As a consequence, it becomes necessary to ensure that the product lines are properly designed and that the models indeed capture the intentions of the modelers. This paper presents a novel methodology to validate the behavior of SPL models by offering simple tools to ``see and compare'' the \emph{actual} behavior of a model with the \emph{expected} one.

To validate models that present quantitative aspects in their behavior, we often use exact or statistical analysis techniques. 
The formal verification of the dynamics of a system via exact techniques provides precise values of the (quantitative) properties being analyzed. These typically require reasoning upon the whole behavior of the system, which might not be feasible for complex models. Indeed, as the possible dynamics of the system increase, these techniques tend to suffer from the well-known state-space explosion problem\rtwo{, rendering them inapplicable when the state-space becomes infinite (see, e.g.,~\cite{ter2021quantitative}).} 
On the other hand, statistical analysis techniques, such as Statistical Model Checking (SMC)~\cite{Agha18}, rely only on limited but statistically relevant samples of executions of a model: simulations. Therefore, statistical analysis techniques can be used to analyze complex dynamical systems\rtwo{, potentially with infinite state spaces,} at the cost that analysis results are not \emph{exact} anymore but are only statistically reliable estimations, e.g., equipped with confidence intervals.

When the overarching behavior of a system is unknown, and it is impossible to make assumptions about its transition structure, the system is referred to as a black-box system. An SMC that analyzes the dynamics of a black-box system without prior knowledge of the system is referred to as a black-box SMC~\cite{younes2005probabilistic}. These simulation-based approaches return numerical estimates, plots, and occasionally counterexamples of the studied properties. However, they typically do not provide behavioral explanations for the results obtained. Without clear explanations, the modeler can only make informed guesses about how to adjust the model to fix unwanted behaviors. For example, let us assume that we consider an SPL model for a family of vending machines, a classic PLE model 
\rev{(see, e.g.,~\cite{DBLP:conf/fm/VandinBLL18,BMS12,CHSLR10,BDV14a,BLLV15a,MPC16,BDVW17})}.
 Let us further assume that we use 
 SMC to study the probability that machines from the family contain dispensers for cappuccino and that we get $0$. Interesting questions \rev{about this analysis} are: 
\begin{itemize}
	\item \emph{What is the reason behind such an extreme value?}, 
	\item \emph{Was the model intended to express this dynamic, or is there a bug?}
\end{itemize}
\rev{In our view, the numeric value $0$ is a black-box analysis result. Meaning that we do not know \emph{why} we got $0$, we do not know if it comes from an issue in the model, nor how to fix it. The core of our proposal is to enrich the analysis results obtained \rtwo{by SMC} to study this query by automatically adding explicit visual information pinpointing any misalignment between the model and the two bullet points above.}~\revsecond{This approach not only facilitates model refinement but also serves as a method for conducting comprehensive testing. Experimenting with diverse settings within the same model structure enables the evaluation of the correctness of the simulated model.}

\rtwo{Our proposal involves in enriching SMC analyses by conducting additional post-processing and analyzing the byproducts of SMC, i.e., log files on the computed simulations, using popular data-driven techniques known as Process Mining (PM). Given that SMC is able to handle models with infinite state-space, our approach can similarly address such scenarios.
}
PM is a process-oriented data-driven technique that analyzes the executions (i.e., traces) of activities generated by information systems, allowing the identification of process patterns, bottlenecks, and other issues in a model~\cite{VanderAalst2016}. Visualizing the flow of activities helps identify opportunities for improvement (e.g., unexpected loops or unexpected dependencies).

This paper, which extends a preliminary work~\cite{10.1007/978-3-031-25383-6_18} where we \rev{sketched} the potential of enriching SMC techniques with PM, presents a white-box technique to enable the evaluation of behavioral aspects of a feature model. The technique leverages the application of process mining techniques on event logs produced by (simulations generated by) statistical model checking. The goal is to provide insights into the system behavior, such as discovering new patterns, identifying bottlenecks, and improving the general model accuracy.  Therefore, integrating SMC with PM paves the way to a more comprehensive understanding of the overall behavior of the model and can help identify issues or suggest actionable improvements to the modeler. To the best of our knowledge, our methodology is the first attempt in \rev{automatically} \emph{explaining} the results of SMC using PM-based approaches. 
In our \rev{preliminary} previous work~\cite{10.1007/978-3-031-25383-6_18}, we \rev{exemplified} the capabilities of analyzing traces generated by SMC with PM \rev{using a preliminary version of our methodology}. However, the identification of issues was entirely delegated to the modeler's visual skills and hence highly subjective. \rev{Furthermore, the preliminary methodology had low accessibility, as the mined model was given using a PM formalism different from the one used to create the original model. This had the additional disadvantage that,} in the presence of a large model, mining the simulations would result in a \rev{complex mined PM model}, making it difficult for the modeler to locate issues. 
In this paper, we overcome these limitations by \rev{fully developing the methodology to} \emph{automatically} discover \rev{and visualize} undesirable behaviors. \rev{Such findings are then shown directly in the model specification itself}. This is accomplished by highlighting the differences between the expected behavior of the model, the model specification, and the actual behavior discovered by mining its simulations. In fact, by highlighting the specific behaviors causing issues in the model, the modeler can make more targeted and effective adjustments to the model to fix those issues. 
\rev{Furthermore, in~\cite{10.1007/978-3-031-25383-6_18}  we did not tailor the SPL domain, but only the cybersecurity one, while we now explicitly target the SPL domain by applying the approach on the feature-oriented language QFLan~\cite{DBLP:journals/tse/BeekLLV20,DBLP:conf/fm/VandinBLL18}.}

\begin{figure}
	\centering
	\includegraphics[width=\textwidth]{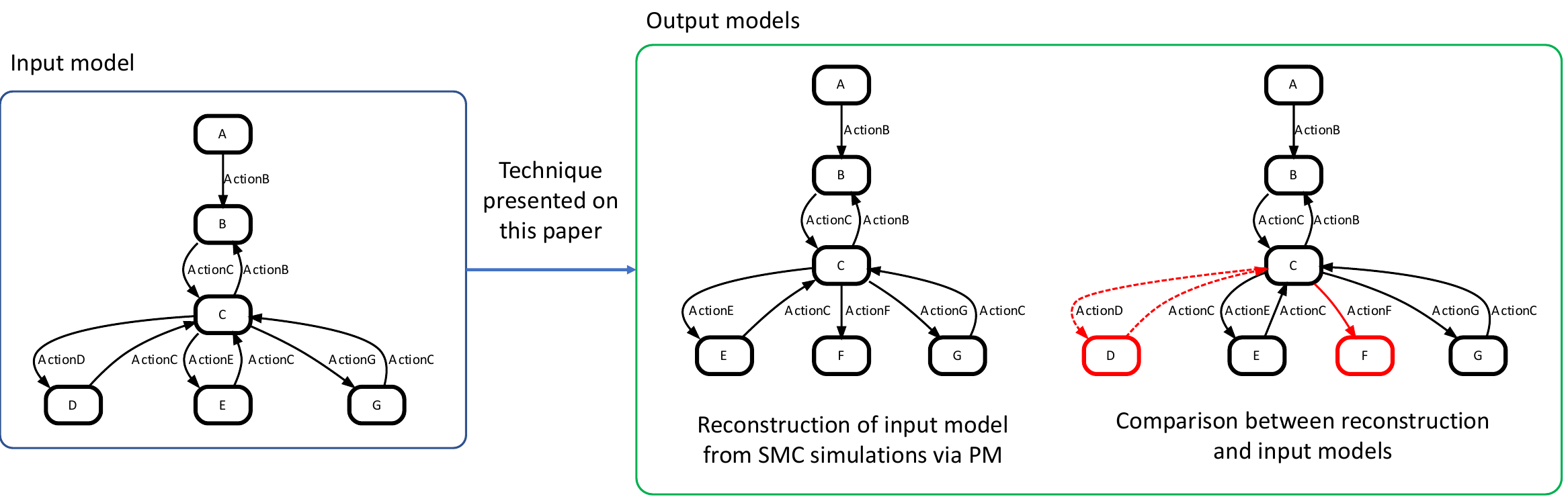}
	\caption{Example of the input and output produced by the technique presented in this paper. The input model is simulated with SMC techniques and corresponding traces are used to synthesize a new model which is then compared to the original one and an easy-to-read output is returned to the modeler emphasizing the differences.}
	\label{fig:abstract_example}
\end{figure}
Figure~\ref{fig:abstract_example} illustrates an abstract example of a model validated using our method. The input model is a representation of an abstract model. 
This abstract model includes different states and actions used to move between those states. In this example, the simulator would start from the node \cf{A} and with the \cf{ActionB} move to the node \cf{B}, and from there can move to another node by choosing the corresponding action. Without our methodology, 
an SMC user would need to validate the input model using only the obtained numerical results (e.g., if interested in ``estimating the probability of reaching a specific node in the model''). With our approach, however, the modeler can inspect the results of the simulation once they are synthesized into new models, thus using the same language (i.e., models-to-model rather than models-to-numbers). The first graph in the ``output models'' box of Figure~\ref{fig:abstract_example} depicts the reconstructed model after applying Process Mining to the simulations. Instead,  the rightmost model depicts the graph obtained by comparing the input model and the reconstructed one. This final representation, in particular, highlights the differences between the input model and the simulated behavior, allowing the modeler to quickly identify issues, such as unexpected or missing behaviors. The exact semantics of the different colors of the edges is explained in Section~\ref{sec: method}.
It is worth mentioning that this method can, in principle, be applied to any discipline where discrete-state simulation models are used, enhancing the capabilities of related modeling and analysis tools. 
On the other hand, the methodology is particularly useful for domains where the complexity of models is high, as in the case of highly-parametric models from PLE. \rtwo{In particular, SMC, and therefore our approach, which post-processes its results, is particularly useful for models with very large or infinite state space models. }

It is important to stress that the methodology depends on the input model, but also on the chosen studied property, as this will drive the simulation process. This is somehow reminiscent of the so-called \emph{CEGAR} (counterexample guided abstraction and refinement~\cite{10.1007/10722167_15}) approaches from qualitative model checking \rev{(and applied only in very limited way to probabilistic settings~\cite{DBLP:conf/cav/HermannsWZ08})}. In classic qualitative model checking, if the studied property does not hold, we get counterexamples of systems' dynamics that showcase executions that falsify the formula. CEGAR involves the use of such counterexamples to refine the model. In our methodology, we proceed similarly, by using mined process models as counterexamples.
\rev{This is orthogonal to static analysis approaches like, e.g., \cite{DBLP:journals/ese/BeekDLMP22}, which aim at identifying issues of the model in general, not tailored to the verification of a single property.}

To validate our methodology we provide positive answers to the following research questions:
\begin{itemize}
\item[RQ1.] \textit{(effectiveness)} Can the developed techniques be \revsecond{employed for a comprehensive evaluation aimed at thoroughly studying the behavior of models and identifying any errors within them}?
To answer this question, we apply our methodology to the feature-oriented quantitative modeling and analysis framework QFLan~\cite{DBLP:journals/tse/BeekLLV20,DBLP:conf/fm/VandinBLL18}. 
We demonstrate the effectiveness of the proposed method by applying it to a model describing a family of beverage Vending machines product line from~\cite{DBLP:conf/fm/VandinBLL18}, a classic case study in PLE. 
Our experiments\revsecond{, in more challenging settings of previous experiments~\cite{DBLP:conf/fm/VandinBLL18},} 
demonstrate the effectiveness of our method: We can automatically \revsecond{conduct a comprehensive evaluation of the behavior of the model}.
\item[RQ2.] \textit{(scalability)} Are the developed techniques scalable to large models considered challenging by the SPL community? To answer this question we apply our methodology to a case study of an Elevator product line from~\cite{DBLP:journals/tse/BeekLLV20}, initially proposed by~\cite{PLATH200153}, which is a well-known case study used to test the scalability in PLE. 
\rtwo{Furthermore, it is worth noting the variant of Vending machines product line considered in this paper is infinite state space.}
Our experiments show that our methodology tends to have a runtime in the same order of magnitude of SMC analysis, and it never exceeds more than 5 times its runtime.
\item[RQ3.] \textit{(multi-domain)} Can the developed techniques generalize to further domains beyond software product lines? To answer this question we consider an additional domain, namely cybersecurity, using the framework RisQFLan~\cite{ter2021quantitative}. It is an incarnation of QFLan to the cybersecurity domain. %
Thanks to this, we successfully validate our approach on an example of a threat model from~\cite{ter2021quantitative}, demonstrating its applicability to different domains.
Indeed, we show how we automatically discovered unwanted and unexpected behaviors. In addition, we also got hints on how to fix such issues by obtaining a refined model that does not show the issues.
\rtwo{Notably, as emphasized in~\cite{ter2021quantitative}, the considered model has infinite state-space, providing additional insights for addressing RQ2.}
\end{itemize}

\rev{All models and replication material for this paper are available at \url{https://doi.org/10.5281/zenodo.8362717}.}

\paragraph{\rev{Synopsis}}The remainder of the paper is structured as follows. Section~\ref{sec:back} introduces necessary background material, as well as the Vending machine as a running example. After this, Section~\ref{sec: method} presents our methodology, while Sections~\ref{subsec: effectiveness}-~\ref{subsec: Multi-domain} validate it on three case studies, answering our research questions. \rev{Section~\ref{sec:related} discusses related works, while} Section~\ref{sec:disc} concludes the paper 
and drafts future works. 









\paragraph{\rev{Further discussion regarding the relationship with~\cite{10.1007/978-3-031-25383-6_18}}}
\rev{ this paper expands upon the preliminary research presented in that work. Specifically, within this paper:
We generalized the approach from the security domain to the SPL one (we added native support for QFLan);
We complete the methodology by computing automatically a \emph{diff} model, given in the original model specification language, to highlight the differences between the reference and mined models;
We evaluate the scalability, effectiveness, and generality of the approach via proper experiments;
We consider a more complex security model; 
We added a related work section and an actual artifact to be used by third parties.}

%% file: background.tex
\section{Background}\label{sec:back}

This section presents the fundamental notions needed throughout the rest of the paper.

\subsection{Modeling product lines with QFLan}
\label{subsec: QFLan tool}


QFLan (Quantitative Feature-Oriented Language) is a feature-oriented language member of the FLan (Feature-Oriented Language) family~\cite{DBLP:conf/fm/VandinBLL18, DBLP:journals/tse/BeekLLV20, DBLP:conf/splc/BeekLLV15}. 
It is based on the principles of concurrent constraint programming and is used to specify the configuration and behavior of product lines mixing procedural and declarative aspects. To achieve this, QFlan employs a constraint store to separate the declarative aspects of the model, e.g., the constraints imposed by a feature diagram, from procedural reconfiguration aspects typical of dynamic SPLs. \rev{In fact, QFLan can deal with aspects of dynamic SPLs such as the staged configurations known from dynamic SPLs~\cite{CHE04,BLLBGS14} (e.g., adding and removing features as well as activating and deactivating features at runtime).
} 
This allows the modeler to express typical constraints from feature models in a declarative manner. These two aspects are unified by the formal semantics of  QFlan. 




QFlan supports quantitative analysis,  via the statistical analyzer \mv{} discussed in Section~\ref{sec:smc}. 
QFLan has been recently recast for the security risk modeling domain, obtaining the language RisQFLan~\cite{ter2021quantitative}. In this paper we consider both QFLan, to show that our methodology can be of interest to the PLE community, and RisQFLan, to show the multi-domain nature of our approach.




\begin{figure}
\centering
\includegraphics[width=.8\textwidth]{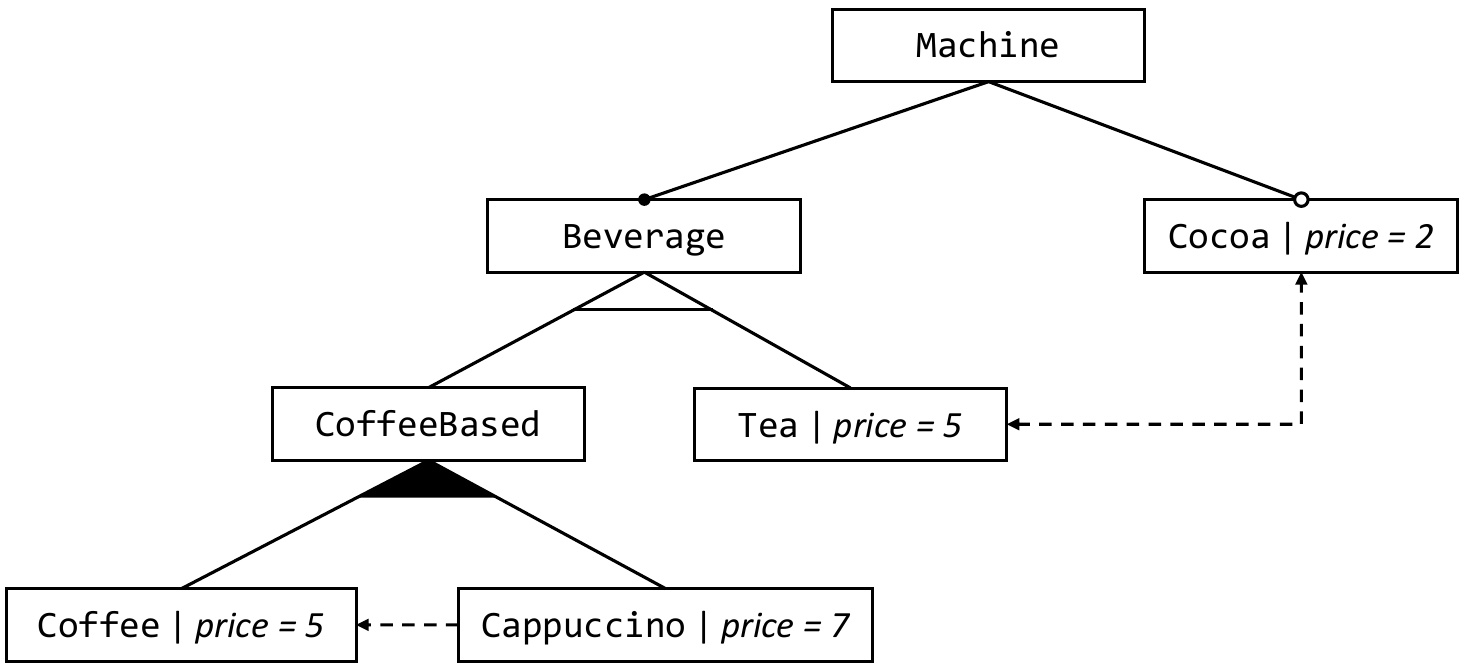}
\caption{Feature model for  hot beverage vending machines, figure adapted from~\cite{DBLP:conf/fm/VandinBLL18}}
\label{fig:qflan_feature_model}
\end{figure}

To ease the presentation of our ideas, we use as a running example a classic case study in PLE presented in~\cite{DBLP:conf/fm/VandinBLL18}. \rev{This is an adaptation of several proposals from the literature, commonly used to present novel methodologies for PLE (e.g.,~\cite{BMS12,CHSLR10,BDV14a,BLLV15a,MPC16,BDVW17}).} 
This is a classic example of a product line of vending machines that offer a selection of tea and coffee-based beverages, including \texttt{Coffee}, \texttt{Cappuccino}, \texttt{Tea}, and \texttt{Chocaccino} (\texttt{Cappuccino} with \texttt{Cocoa}). 
\rev{Each product of this family is a concrete vending machine with an \emph{admissible} subset of beverages.}
Figure~\ref{fig:qflan_feature_model} depicts graphically the \emph{feature model} of our example. It describes the structural constraints among the \emph{features} that may be present or not in valid instances of vending machines. Each node represents a feature, and the edges between nodes represent constraints that define the admissible combinations of features. QFLan has two types of features: concrete and abstract. The four leaves represent concrete features. These can be \emph{installed} or \emph{uninstalled} explicitly. Instead,  abstract features are internal nodes. These are not explicitly (un)installed, rather they are implicitly added or removed when their children nodes are. Abstract features are mainly used to group related features. The root node represents a complete product, which in this case is a specific vending machine.  The feature diagram imposes that concrete instances of vending machines may or may not contain the optional feature \cf{Cocoa} (empty circle over it), but its presence excludes the possibility of serving teas (and vice-versa, the dashed line connecting the two). A concrete vending machine is required to have a beverage feature, which can be either coffee-based or \cf{Tea} (the edges with an empty triangle connecting the three features).
Coffee-based beverages can be \cf{Coffee} or \cf{Cappuccino}, with the \emph{cross-tree constraint} that the latter need \cf{Coffee} (the dashed arrow from \cf{Cappuccino} to \cf{Coffee}).
QFLan models are actually given using a textual representation in a specific domain-specific language. The full model specification can be found in~\cite{DBLP:conf/fm/VandinBLL18}. Here we provide the minimum information to make the paper self-contained. 

\begin{figure}
	\centering
	\begin{lstlisting}
		begin action constraints 
		// Machines serving Cappuccino and Cocoa dispensers can serve chocaccino
		do(chocaccino) -> (has(Cappuccino) and has(Cocoa)) 
		end action constraints
	\end{lstlisting}  
	\caption{Action constraints}
	\label{fig:qflan_Action_constraints}
\end{figure}

\begin{figure}[t]
	\centering
	\begin{lstlisting}
		begin quantitative constraints
		//The price of generated products must never exceed 10 (or 15)
		//{ price(Machine) <= 10 }
		{ price(Machine) <= 15 }
		end quantitative constraints 
	\end{lstlisting}  
	\caption{Quantitative constraints}
	\label{fig:qflan_quan_constraints}
\end{figure}

We note that \cf{Chocaccino} does not belong to the feature diagram in Figure~\ref{fig:qflan_feature_model}. This is because it is not a feature to be (un)installed. Instead, it is a sort of implicit feature available when the machine can serve both \cf{Cappuccino} and \cf{Cocoa}. This is encoded by the \emph{action constraint} in  Figure~\ref{fig:qflan_Action_constraints}. 
QFlan allows to consider further classes of constraints, including quantitative ones. E.g., the price of a sold vending machine could be limited to a maximum cost computed by summing the cost of currently installed concrete features (see Figure~\ref{fig:qflan_quan_constraints}). As we will see in Section~\ref{subsec: effectiveness}, these constraints considerably impact the dynamics of the model.

\begin{figure}
\centering
\begin{lstlisting}
    begin processes diagram
    begin process dynamics
        states = factory , deposit , operating , prepareCoffee , prepareCappuccino, prepareTea , prepareChocaccino
        transitions = 
        // Factory state
        factory -(replace(Coffee,Tea),20)->factory,
        factory -(install(Cocoa),10)->factory,
        factory -(install(Cappuccino),10)->factory,
        factory -(uninstall(Cappuccino),10)->factory,
        factory -(sell,1,{sold=1})-> deposit,(*\label{lst:linesold}*)
        //Deposit state
        deposit -(install(Cappuccino),2.0)->deposit,
        deposit -(uninstall(Cappuccino),2.0)->deposit,
        deposit -(install(Cocoa),2.0)->deposit,
        deposit -(uninstall(Cocoa),2.0)->deposit,
        deposit -(deploy,2,{deploys=deploys+1})-> operating ,(*\label{lst:linedeploy}*)
        // Operating state 
        // Serving Coffee
        operating -(Coffee,3)-> prepareCoffee,
        prepareCoffee -(serveCoffee,1) -> operating,
        // Serving Cappuccino
        operating -(Cappuccino,3)-> prepareCappuccino,
        prepareCappuccino -(serveCappuccino,1) -> operating,
        // Serving Chocaccino
        operating -(chocaccino,2)-> prepareChocaccino,
        prepareChocaccino -(serveChocaccino,1) -> operating,		
        // Serving Tea
        // ...
        operating -(reconfigure,1) -> deposit
    end process
    end processes diagram
    
    begin init
        installedFeatures = { Coffee }
        initialProcesses = dynamics
    end init
\end{lstlisting}  
\caption{Probabilistic process of the model in QFLan 
}
\label{fig:qflan_transition_system_VM}
\end{figure}

\begin{figure}
	\centering
	\begin{lstlisting}
    begin actions
        sell deploy reconfigure
        serveCoffee serveCappuccino serveChocaccino serveTea chocaccino
    end actions
        
    begin variables 
        sold = 0    deploys = 0
    end variables
	\end{lstlisting}  
	\caption{Actions and variables}
	\label{fig:qflan_actions}
\end{figure}


In addition to the discussed constraints, the declarative part, QFLan models come with a procedural part. This specifies the dynamic behavior of the model. 
Specifically, Figure~\ref{fig:qflan_transition_system_VM} lists the probabilistic process of the Vending machine. This includes different states and transitions among them. Transitions must be labeled with weights, used to compute the probability of executing a transition and actions. Actions can be feature names, which signal the use of installed features, or custom actions (listed in Figure~\ref{fig:qflan_actions}). We also note that transitions might be further labeled with \emph{side-effects} which change the value of variables (see, e.g\rev{.}, line~\ref{lst:linesold} where we set variable \cf{sold} to 1).  Variables are declared in the \cf{variables} block, see Figure~\ref{fig:qflan_actions}\rtwo{, and implicitly might have an infinite domain. Notably, in~\cite{ter2021quantitative} we show that the fact that variables can take infinite values easily leads to a model with infinite state spaces. As discussed in~\cite{ter2021quantitative}, this limits the application of exact analysis techniques, and required to consider SMC.}
As we can see from the \cf{init} block, the first state is the \cf{factory}, and the machine is initialized with only the \texttt{Coffee}. According to the transitions of state \cf{factory}, e.g., \cf{Coffee} can be replaced with \texttt{Tea}, and the other two beverages can be installed. When the dispenser is sold (line~\ref{lst:linesold}), it moves to state \cf{deposit}. \rtwo{Notably, every time a \cf{deploy} action is performed, the variable \cf{deploys} is increased by one, leading potentially to an infinite state space.}
The end user can customize the dispenser by installing or uninstalling one of the beverages. When ready, the dispenser is deployed (line~\ref{lst:linedeploy}) moving into \cf{operating} state where the installed beverage can be served. Finally, the dispenser can be sent back to the \cf{deposit}. 


\begin{figure}
\centering
\includegraphics[width=0.95\textwidth]{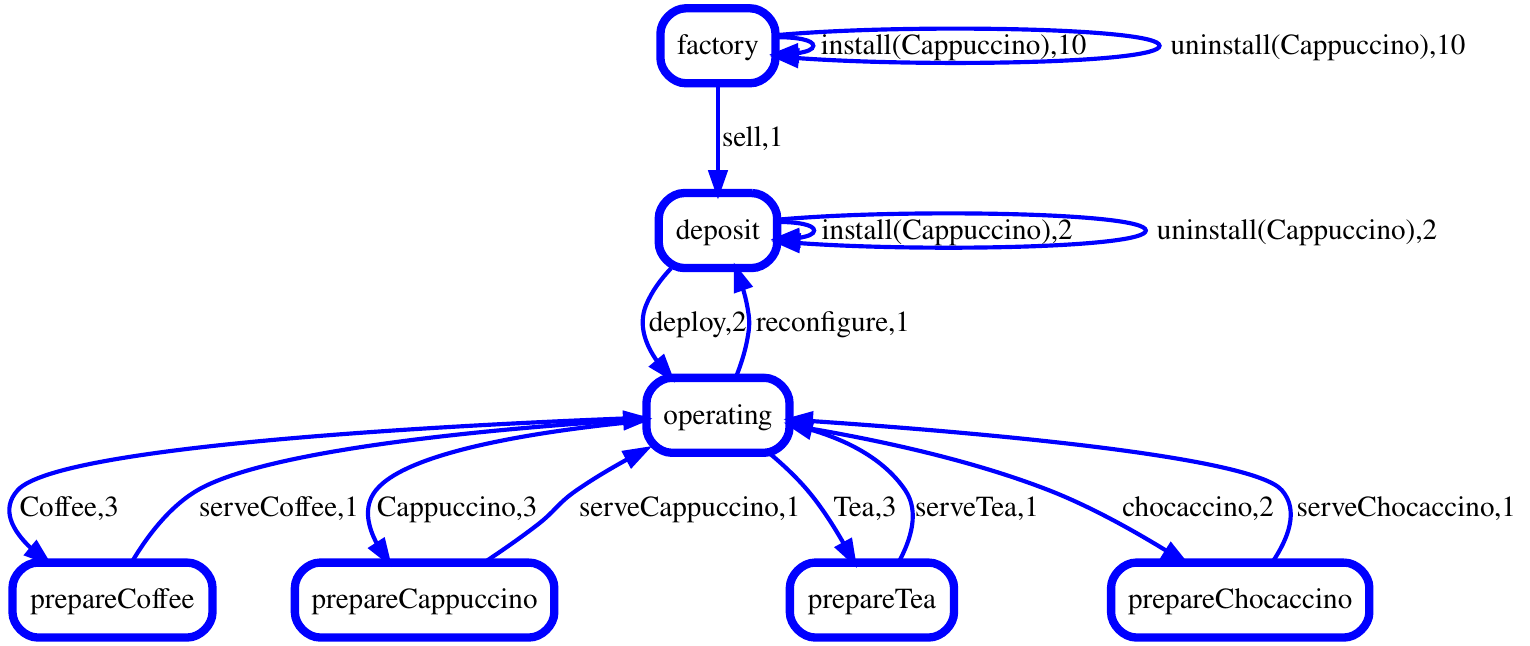}
\caption{Graphical representation of procedural part of the QFLan model of the vending machine. Automatically generated in dot format by QFLan (edited by hand to improve readability). 
}
\label{fig:qflan_model_behavior}
\end{figure}

This procedural specification can be graphically depicted as in Figure~\ref{fig:qflan_model_behavior}, automatically generated by QFLan, edited by hand to improve readability, \rev{e}.g., some \emph{loop} edges have been removed. 
For instance, the graph is missing \cf{replace(Coffee,Tea)} and \cf{install/uninstall(Cocoa)} in \cf{factory} and in \cf{deposit}, despite these being present in the actual model. 

\subsection{Black-box analysis of simulation models with Statistical Model Checking}
\label{sec:smc}
QFLan models can be analyzed by 
black-box SMC~\cite{Agha18} using the tool MultiVeStA which can be plugged into existing simulators~\cite{mvjedc2022}. Given a quantitative property of interest, e.g.\rev{,} the probability of installing a feature or the average price of sold vending machines, \mv{}  performs \textit{enough} probabilistic simulations of the model to obtain statistically reliable estimations of the property. 
Black-box SMC is a simulation-based approach where only probabilistic simulations of the model are performed, with no assumption about the overarching behavior of the model. \mv{} is an example of a black-box SMC tool that can perform statistical analyses over multiple properties simultaneously and is highly scalable~\cite{mvjedc2022,DBLP:conf/hpcs/PianiniSV14}. The tool enables the user to query for one or more properties of the model they want to estimate and returns the estimation of those properties within confidence intervals. 
For instance, if $X$ is a random variable giving the price of sold bikes in a simulation, then \mv{} will estimate its expected value $E[X]$ as the mean $\overline{x}$ of $n$ independent simulations, with $n$ large enough but minimal, to build a $(1-\alpha)*100\%$ of width at most $\delta$ centered on $\overline{x}$. 
In other words, MultiVeStA guarantees that $E[X]$ belongs to the interval $[\overline{x} - \sfrac{\delta\!}{2}, \overline{x} + \sfrac{\delta\!}{2}]$ with statistical confidence of $(1-\alpha)\cdot100\%$. The generation of new samples ends when the confidence interval size is less than or equal to $\delta$ ($\alpha$ and $\delta$ are user-specified parameters). MultiVeStA has been successfully applied to various domains, including security risk modeling~\cite{ter2021quantitative}, economic agent-based models~\cite{mvjedc2022}, highly-configurable systems\rev{~}\cite{DBLP:journals/tse/BeekLLV20,DBLP:conf/fm/VandinBLL18}, public transportation systems~\cite{gilmore2014analysis,ciancia2016tool}, 
lending pools in decentralized finance~\cite{DBLP:conf/isola/BartolettiCJLMV22},
business process modeling~\cite{DBLP:journals/jss/CorradiniFPRTV21}, robotic scenarios with planning capabilities~\cite{belzner2014reasoning}, and crowd steering scenarios~\cite{DBLP:conf/hpcs/PianiniSV14}. 
 Classic qualitative model checking, where properties are either satisfied or not, is able to provide counterexamples whenever a checked property does not hold. This is an example of system execution that falsifies the formula. Unfortunately, counterexamples are not common in quantitative variants of model checking, and even less in SMC.
 A common downside to most SMC approaches is that it does not provide behavioral explanations about why a property is estimated to a given value. This is what we want to solve with our methodology. 

\rtwo{We remark that QFLan allows to perform analyses on the overall family, and not of single products. For example, we can study the probability of having a given feature installed in products of a family, the average cost of products of a family, etc. 
QFlan 
does not directly allow analyses of properties specific to an individual product within that family. However, studying properties of a single product could be accomplished by constraining the probabilistic process 
to focus solely on that product rather than the entire product range within the family. Alternatively, defining the set of installed features in the \cf{init} block of the QFLan model in a manner that permits the analysis of a single product is another approach.}

\begin{figure}
\centering
\includegraphics[width=1\textwidth]{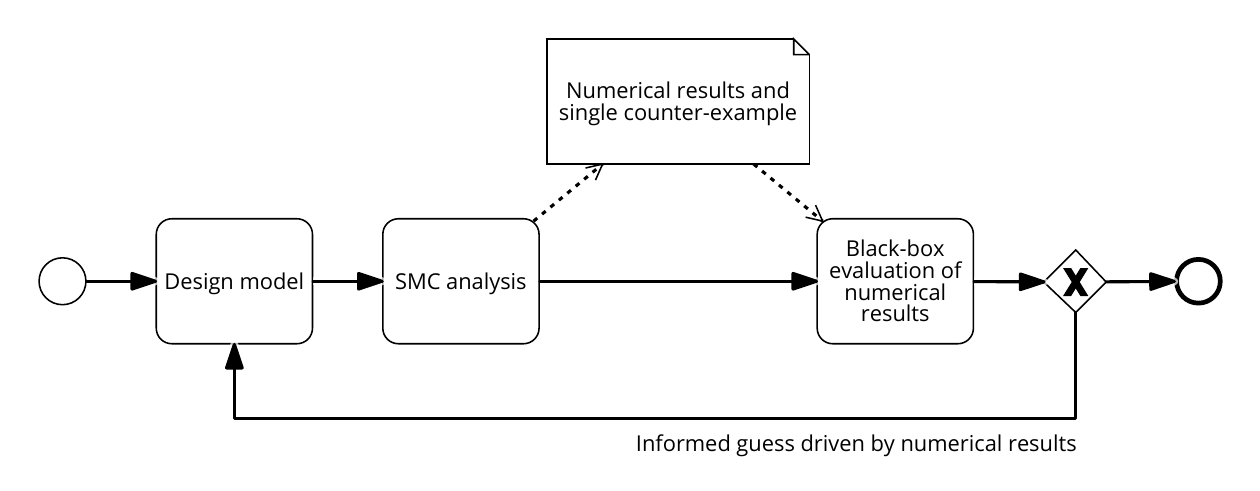}
\caption{Sketch of SMC-based black-box validation}
\label{fig:diagram0}
\end{figure}

\paragraph{State-of-the-art SMC-based validation process}
Figure~\ref{fig:diagram0} illustrates the state-of-the-art process adopted in a traditional SMC setting. 
The process begins with the modeler, who creates the model and then instructs the SMC to estimate properties of interest for the system being modeled. 
SMC returns estimations of the properties 
without providing any additional information on why the results were obtained. SMC  might provide single counter-examples, e.g.\rev{,} an interesting simulation, but to the best of our knowledge, there are no SMC approaches that try to combine simulations to obtain a representation of the dynamics that led to and explain a given estimation. 
In case the estimates are inconsistent with the expectations of the modeler, 
the modeler must make an informed guess on how to modify and correct the model. We call this process \emph{SMC-guided black-box validation}. This is because any decisions to alter the model are made in a black-box manner without knowing the reasons behind the results of SMC. 
From this discussion,  it emerges the need for a methodology like ours that aims at identifying unwanted behaviors and highlighting their origin.

\subsection{Synthesis of models from their executions using Process Mining}
\label{subsec:process-mining}

Process Mining (PM) is an interdisciplinary field that seeks to extract insights from the actual executions of a process by bridging the gap between data science and process science~\cite{VanderAalst2016}. The main activities of PM include discovery, enhancement, and conformance checking. Discovery involves identifying an abstract representation of the executed process by combining all the observed instances into a single model. Enhancement enriches the model with additional information, such as the frequency of executed activities or paths. Finally, conformance checking assesses the extent to which a normative model deviates from actual executions. 

In this work, we are interested in the discovery and enhancement tasks of PM. Specifically, we aim to use execution traces (simulations) obtained from SMC analyses to synthesize new models that capture the behavior of the model as observed in the simulations\rtwo{, even for models with infinite states}. To accomplish this goal we employ the Heuristics Miner (HM) algorithm~\cite{weijters2006process, VanderAalst2016}. 
\begin{figure}[t]
\centering
\includegraphics[width=0.7\textwidth]{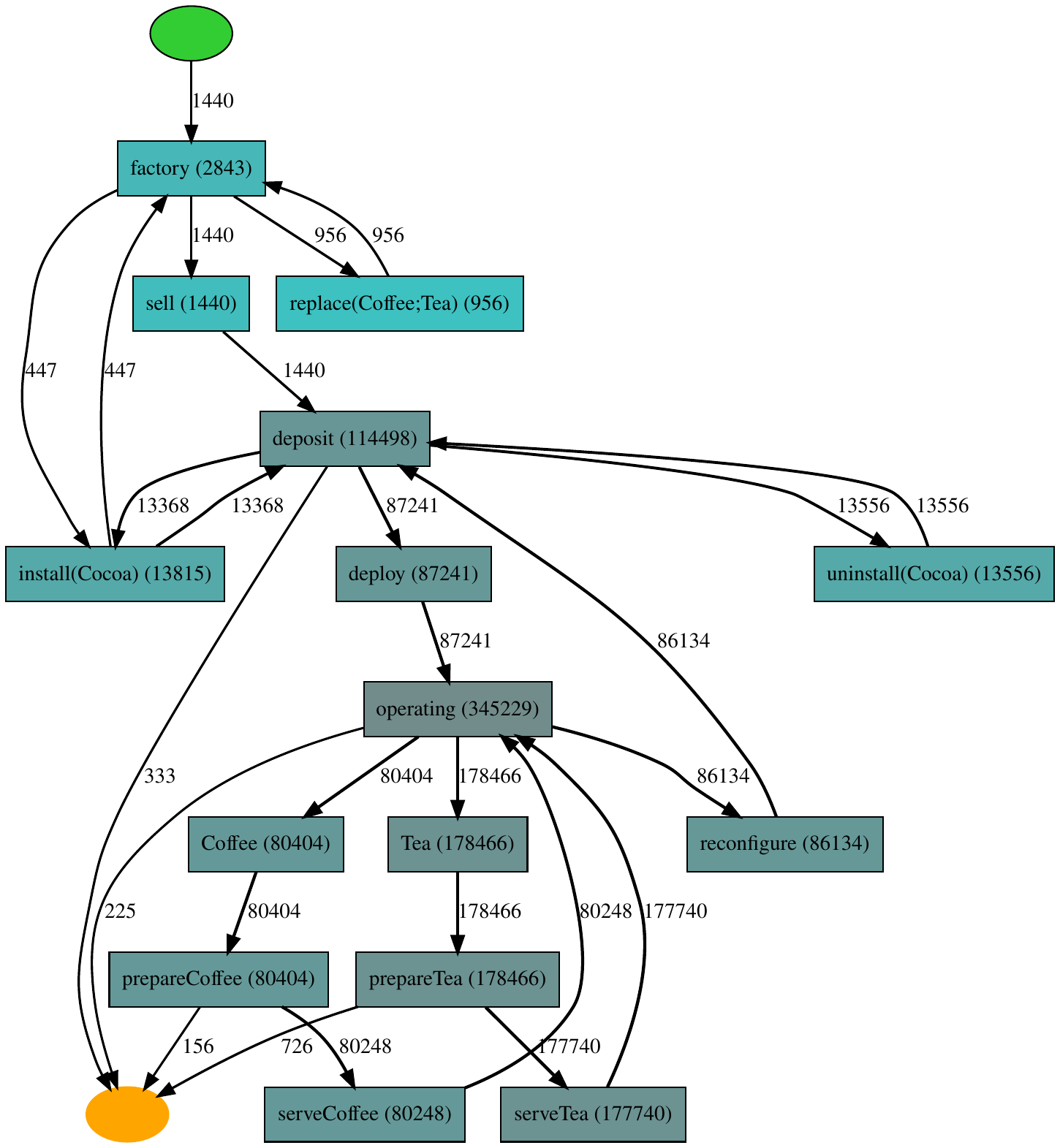}
\caption{Heuristic Net simulation Vending machine model}
\label{fig:PM VM}
\end{figure}
The HM algorithm can provide an accurate and comprehensive understanding of complex process dependencies, facilitating the alignment (the comparison) of the generated and the expected behavior of a model. In addition, the HM algorithm allows the user to adjust some parameters to control the trade-off between model fitness and the inclusion of infrequent paths, such as the noise threshold and dependency threshold parameters. Adjusting these parameters increases the likelihood of including infrequent paths in the discovered process model. In this regard, we follow a conservative approach that preserves any behavior observed during the simulations.

Figure~\ref{fig:PM VM} depicts a Heuristic Net (HN) obtained by applying HM on the simulations of our running example (for the constraint of the maximum price of sold machines set to  10). 
As we can see, an HN consists of nodes connected by edges labeled with frequencies. In HN jargon, states are known as \emph{activities}. 
In our methodology, QFLan activities and states get flattened in the same notion of activity in an HN.
As we can see, the HN has two additional states, the green and red circles, that represent the \emph{start} and \emph{end} state, respectively, of the mined process model. 

%% file: method.tex
\section{Method }
\label{sec: method}

\begin{figure}[t]
	\centering
	\vspace{-0.1cm}	
	\includegraphics[width=0.93\textwidth]{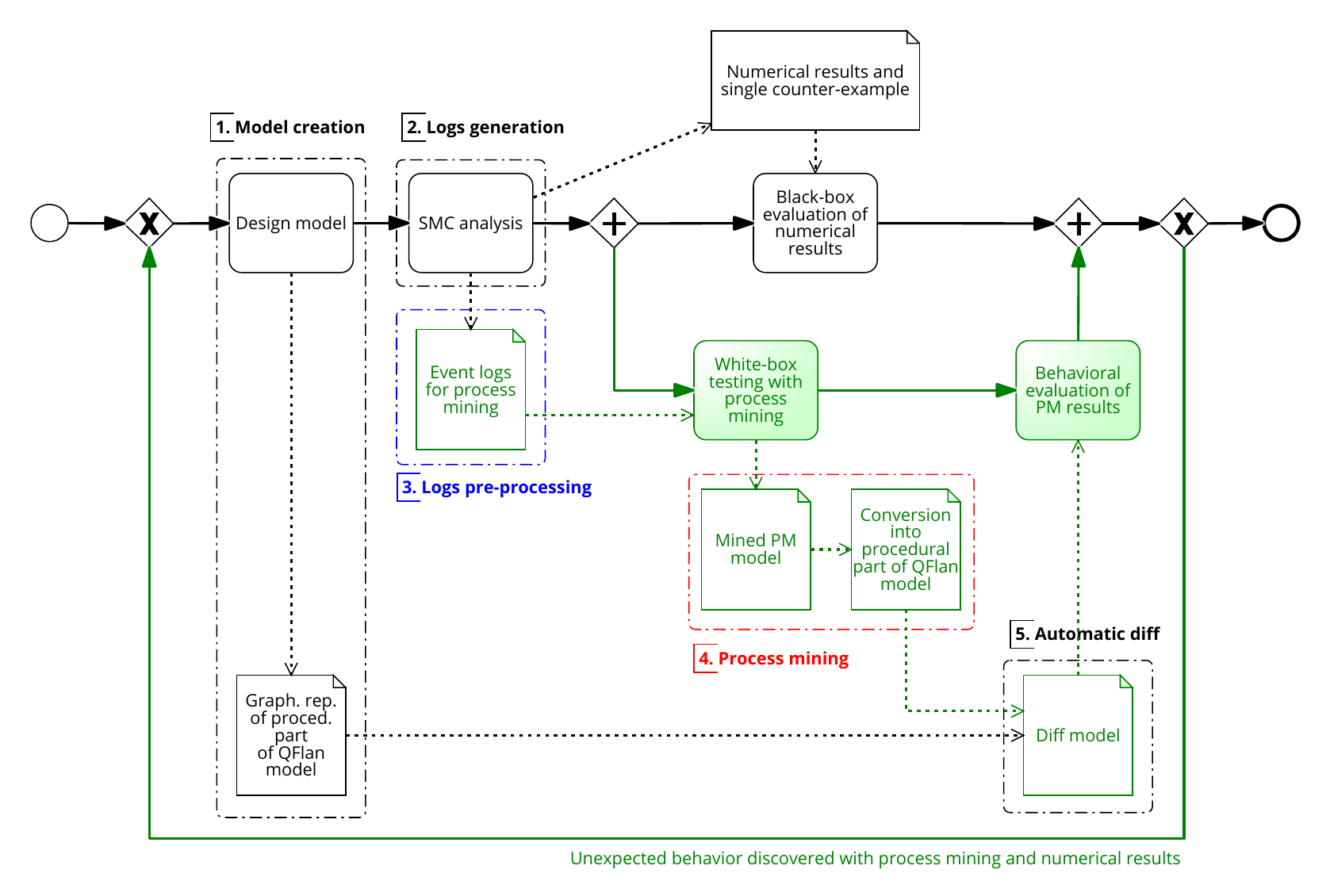}
	\vspace{-0.4cm}
	\caption{The new methodology presented in this paper, combining SMC and PM for \textit{automatic} white-box behavioral validation. Green activities/data-objects/sequences identify novel aspects introduced in this paper. The main steps of the procedure are numbered. 
	}
	\label{fig:diagram}
\end{figure}

 
We propose a method that enriches SMC techniques with PM techniques to overcome the limitations of the classic SMC black-box validation seen in Section~\ref{sec:smc}. \rev{In our previous study~\cite{10.1007/978-3-031-25383-6_18}, we have demonstrated the effectiveness of combining SMC analyses with PM techniques in identifying undesired behaviors in formal models. The integration of PM allows us to incorporate a white-box analysis of the model's behavior by leveraging mining techniques on simulated model executions. However, our earlier work was preliminary and not automated: it relied primarily on the modeler to visually identify undesirable behaviors as depicted in the PM output. Furthermore, the PM output was given in a formalism different from the one model specification language, making our approach less accessible. To solve these issues, in this current study, we go beyond a simple discovery algorithm applied to event logs. Instead, we establish an integration between SMC and PM techniques through a graphical component given in the model specification language. This component takes SMC logs as input, applies PM techniques to analyze them, and then visualizes the results using the original model specification language. This new approach 
	facilitates the automatic discovery of missing or undesirable behaviors in the model.} 

Figure~\ref{fig:diagram} illustrates the proposed methodology. It enhances SMC with 
automatic white-box behavioral model validation. 
\rev{The presented methodology incorporates several enhancements compared to our previous work. To begin with, prior to applying the discovery algorithm to SMC logs, we now conduct a pre-processing step, which is described in Section~\ref{subsec: Pre-processing}. We then mine a PM model and convert it into a graph that represents the procedural aspect of the QFLan model. 
Subsequently, we compare this mined model with the original graphical representation of the procedural part of the QFLan model. 
The comparison yields a \emph{diff} model, which highlights any disparities in behavior between the formal model designed by the modeler and the actually simulated model. In our previous work, we demonstrated the potential of our approach solely by applying a discovery algorithm to unprocessed SMC logs in order to extract a PM model. An example of the outcome of our previous work is the PM model depicted in Figure~\ref{fig:PM VM}, which was obtained by mining the SMC logs of our running example with the maximum price of sold machine set to 10 using the HN algorithm. \revsecond{In this process model, it is not possible to visually evaluate whether those transitions represent the entire behavior of the model or merely a subgroup of the transitions.}} 
\revsecond{However, as discussed in Section~\ref{results: Effectiveness}, using our current methodology, we can thoroughly evaluate whether the model behaves correctly by identifying transitions that are absent in the simulated model when altering the quantitative constraints within the model.}

\rev{More precisely, }our methodology 
consists of five steps, numbered in Figure~\ref{fig:diagram}. 
In step \emph{1. Model creation}, the modeler creates a model and the graphical representation of its procedural part using a model specification language. For example, QFLan, where the graphical representation is generated automatically from the model description. 
Then, in \emph{2.  Logs generation} we use an SMC tool to run simulations of the model to study a given property. Information on each simulation is stored as a log of events (an \emph{event log}), containing, e.g., time stamps, actions executed, etc. In this paper, we consider the SMC tool \mv{} which has been extended with log generation capabilities. 
Once the simulated event logs are obtained, we pre-process them in step \emph{3. Logs pre-processing}.
 In step \emph{4. Process mining}, we apply PM techniques on these logs to discover the process model describing the behavior of the model as observed in the simulation. In the figure, we call this the \emph{mined PM model}. We then post-process the PM result to convert it into (the procedural part of) a QFLan model. 
 Finally, in step \emph{5. Automatic diff}, we compare the graphical representation of the original model with the one discovered in our step 4.  The result is a graphical representation, in terms of the source modeling language, highlighting the differences between the expected behavior of the model with the real one. We call this the \emph{diff model}. As we will see in our experimental section, the diff model can explain the results obtained by SMC and suggest fixes if necessary.  
 All steps of our methodology (apart from the initial model design) are fully automated.
 
 We refer to this methodology as \emph{white-box behavioral model validation} because, thanks to the union of SMC and PM, we can access the internal workings of the system by shedding light on its actual behavior. Therefore, the modeler can now rely on more than just an informed guess to fix the model, the diff model. 
In the remaining parts of this section, we describe how we implement each of these steps. 


\subsection{Model creation}
\label{subsec: model creation}
The first step of our methodology starts with creating the (QFLan) model where the modeler defines all the components of the system, e.g., features, variables, a list of constraints, and its procedural part. QFLan will then automatically generate 
a \emph{graphical representation of the procedural part of the QFLan model}. 
The one for our Vending machines running example is sketched in Figure~\ref{fig:qflan_model_behavior}.
This is what in Figure~\ref{fig:abstract_example} we call the ``input model''. 
%


\subsection{Logs generation}
\label{subsec: SMC analysis and logs generation}
In this step, the modeler chooses the properties of interest for the model and evaluates them using \mv{}~\cite{mvjedc2022,sebastio2013multivesta}. For example, we might be interested in the average price of the sold vending machines. \mv{} will instruct the simulator to run the required simulations, saving information on them as event logs.  

\mv{} has a clear interface to plug into new simulators only involving three functionalities: \emph{reset to perform a new simulation}, \emph{perform one step of simulation}, \emph{evaluate an observation in the current simulation state}~\cite{mvjedc2022,sebastio2013multivesta}.
To enable log generation, we added two new functionalities to the interface of \mv{}: \emph{create an empty log file}, invoked once per SMC analysis, and  \emph{add row to log}, invoked whenever an event (of interest) is to be recorded. These functionalities have to be implemented whenever integrating \mv{} to a new simulator. When implementing the latter functionality, the modeler might decide to record all events, i.e.\rev{,} all simulation states, or only selected events of interest.
In QFLan,  we add a row whenever we perform one step of a simulation. The recorded information includes the incremental counter of steps (i.e., the time stamp), the unique random seed used by the current simulation (used as case ID, the unique identifier of the simulation/case), the executed action (i.e., the activity), the target state of the executed transition, and any relevant additional information (features currently installed, and values of variables). All information is stored in separate columns and saved in a CSV file. 

\subsection{Logs Pre-processing}
\label{subsec: Pre-processing}
In this step, we pre-process the event logs stored before applying PM techniques. The pre-processing consists of merging the columns that record the target states and the actions used to move from one state to another. This means that states and actions will be treated as activities when we apply the PM discovery algorithm. In Section~\ref{subsec:process-mining}, we showed an example of the HN mined from event logs generated from our Vending machines running example, and pre-processed as discussed here. 
Connected to the merging of the two columns, in order to preserve the correct order to avoid losing information about the transition that executed a given action, we change the name of the actions by adding the names of the origin and target states. Renaming an action is essential because the same action can appear in different transitions across different states. Without such renaming, we would lo\rev{o}se information on the actual executed process. For instance, in the ``input model'' of Figure~\ref{fig:abstract_example}, when choosing \cf{ActionC} to move from  \cf{B} to \cf{C}, we change the name of the action from \cf{ActionC} to \cf{ActionC\_B\_C}. Instead, in the case of execution of the action to move from \cf{E} to \cf{C}, we would get  \cf{ActionC\_E\_C}.

\subsection{Process mining}
\label{subsec: process_mining_step}
We now mine the pre-processed event logs using the Heuristic Miner (HM) algorithm~\cite{weijters2006process, VanderAalst2016} discussed in Section~\ref{subsec:process-mining}. We use the library PM4PY\rev{\footnote{We use the parameters, i.e., \cf{dependency\_threshold=0.5}, \cf{and\_threshold=0.65}, \cf{loop\_two\_threshold=0.5} and \cf{dfg\_pre\_cleaning\_noise\_thresh = 0}. See \href{https://pm4py.fit.fraunhofer.de/documentation}{https://pm4py.fit.fraunhofer.de/documentation}}}~\cite{DBLP:journals/corr/abs-1905-06169}, a versatile Python library that can help in using  different PM algorithms. 
Once the Mined PM  model 
is discovered, 
we then parse it to extract edges and nodes and use them to convert it from a \emph{PM model} (i.e.\rev{,} a Heuristic net mentioned in Section~\ref{subsec:process-mining}) into a \emph{mined QFLan model} (actually, the procedural part of a QFLan model). In Figure~\ref{fig:abstract_example}, this corresponds to the left graph of the ``output model''. 
In this process, we revert the names of the actions to their original ones. This helps in comparing the original QFLan model with the mined one. 

\subsection{Automatic diff}
\label{subsec: Automatic diff}
The last step of our methodology starts by parsing the graphical representations of (the procedural part of) the original QFLan model from step 1 and of the mined one from step 4. 
This allows us to compare the two models, and create a \emph{diff model} that highlights the existing differences. 
The diff model is built as follows: it includes the edges and nodes present in both models, without highlighting them (i.e.\rev{,} it uses the same color and style as in QFLan). 
Then, we add all edges and nodes that appear in only one of the models, this time highlighting them in red. 
The rightmost graph of the ``output model'' of Figure~\ref{fig:abstract_example} depicts the diff model for the other two graphs in the figure. 
We can see that the red dashed edges (e.g.\rev{,} from node \cf{C} to node \cf{D}) denote edges present in the original model, but missing in the mined one. 
Vice versa, red continuous edges (e.g.\rev{,} from node \cf{C} to node \cf{F}) denote edges not present in the original model, but present in the mined one. 

Therefore, dashed red edges denote transitions that the simulator has never taken, implying that the formal model includes some constraints that might \emph{always prevent} those transitions.
\rev{ } We remark that this information might also help the modeler when testing the effect of new constraints, by modifying the model and observing the result of our methodology. 
 As demonstrated in the experiments in Section~\ref{subsec: effectiveness}, in some cases, the modeler could be interested in intentionally varying some constraints, such as quantitative constraints, to understand their effective impact on the behavior of the model. 
 Indeed, by using a classic SMC black-box validation approach, it might not be possible to detect the impact of a constraint on the obtained numerical values that only summarize the estimations of some properties of interest. \revsecond{In \ref{sec:appendix}, we provide an example of an analysis conducted on the Elevator model, which is discussed in Section~\ref{subsec: rq2 scalability}. This instance emphasizes how a minor typo in the probabilistic process of the model can cause unexpected behavior which impacts the model evaluation via SMC, and how our methodology can spot this issue.} 

Instead, continuous red edges, such as the edge between nodes \cf{C} and \cf{F}, represent transitions present only in the mined model. In QFLan, this can only happen in case the simulator gets stuck in a \emph{deadlock state}, i.e., it is in a state where there would be transitions to execute, but they are all disabled by the constraints.
In other domains, instead, we might in principle have further classes of continuous red edges. Alternatively, there might be cases in which one implements the \emph{add row to log} functionality of \mv{} such that it adds extra rows under predetermined conditions. 
The presence of continuous red edges in the diff model might signal errors that could compromise the validity of the results obtained by SMC. This is exemplified in detail in Section~\ref{subsec: Multi-domain}, where we consider the security domain. There, we demonstrate that a property has a low probability (probability of an attacker succeeding in a robbery) just because of bugs in the model. This bug is identified and fixed thanks to our methodology.
 This gives new opportunities to improve the model that were not possible with the classic SMC black-box validation method.


In the following sections, we apply the presented methodology to answer our research questions. We consider the running example (effectiveness), a parametric SPL model (scalability),  and a security model (multi-domain). 



%% file: validation.tex


\section{Experimental evaluation: RQ1 Effectiveness}
\label{subsec: effectiveness}
To answer RQ1, we consider our running example from  Section~\ref{subsec: QFLan tool}: an SPL model of vending machines from~\cite{DBLP:conf/fm/VandinBLL18}. The goal of this section is to illustrate the effectiveness of our methodology.

\subsection{Domain description - Coffee Vending Machine}

Besides the hierarchical, cross-tree, and action constraints, QFLan allows for another essential class of constraints: quantitative constraints exemplified in Figure~\ref{fig:qflan_quan_constraints} for the considered case study. In this case, the quantitative constraints specify the maximum \cf{price} that a vending machine might have. In particular, Figure~\ref{fig:qflan_quan_constraints} imposes that no machine will be produced that costs more than 10 or 15 euros (depending on which of the two constraints is used). Changing such constraints changes the behavior of the model. This is because some beverages cannot be installed in the dispenser if this is too low.

For instance, in~\cite{DBLP:conf/fm/VandinBLL18} the authors focus on \emph{sold machines}, i.e., those obtained after executing action \cf{sell} from state \cf{factory} in Figure~\ref{fig:qflan_model_behavior}. For such machines, the authors show that if the maximum price is 10, then the probability of a sold machine having a \texttt{Cappuccino} dispenser is zero. This is because the hierarchical constraints in Figure~\ref{fig:qflan_feature_model} 
impose that, in order to have \texttt{Cappuccino}, a machine must serve \texttt{Coffee}. The latter costs 5, while \texttt{Cappuccino} costs 7 (Figure~\ref{fig:qflan_feature_model}), for a total cost of 12 that would falsify the constraint on price. 
It is also shown that the probability of installing a \texttt{Cappuccino} increases if the constraint on the price is relaxed to a maximum of 15.

Our experiments follow a strategy similar to that of~\cite{DBLP:conf/fm/VandinBLL18}, but considering a more challenging setting: we add an \cf{uninstall(Cappuccino)} transition in the \cf{factory} state. This has the effect that a probability 0 of having \texttt{Cappuccino} after \cf{sell} action (Line~\ref{lst:linesold} of Figure~\ref{fig:qflan_transition_system_VM}) does not guarantee that \texttt{Cappuccino} has not been installed at all, because it might have been uninstalled. Therefore, the black box analysis in~\cite{DBLP:conf/fm/VandinBLL18} would not guarantee that the constraint has never been violated before selling the machine. 
Instead, we show here that our white-box approach can guarantee this.

\begin{figure}[t]
	\centering
    \begin{lstlisting}
    //Query 1
    begin analysis
        query = when sold == 1 :
        {price(Machine) [delta=0.5], Coffee, Tea, Cappuccino, Cocoa}
        default delta=0.05    alpha = 0.05    parallelism = 1
        logs ="log_name_sold.csv"
    end analysis 

    //Query 2
    begin analysis
        query = eval from 1 to 500 by 1 :
        {price(Machine) [delta=0.5], Coffee, Tea, Cappuccino, Cocoa}
        default delta=0.05    alpha = 0.05    parallelism = 1
        logs ="log_name_steps.csv"
    end analysis 
    \end{lstlisting} 
\caption{Two \mv{} queries to analyse the Vending machine model}
     \label{fig:qflan_query}
    \end{figure}

\subsection{Experiments}
We run two experiments for two configurations obtained by setting to 10 and 15, respectively,  the maximum accepted price. We study the probability of having \cf{Cappuccino} installed right after the \cf{sell} action. We also study the average price of sold machines, as well as the probability of having \cf{Tea}, \cf{Coffee}, and \cf{Cocoa}. 
To run the experiments, we invoke \mv{} using the QFLan GUI.  
We consider the two queries in Figure~\ref{fig:qflan_query}.
Query 1  instructs \mv{} to evaluate the properties on the first simulation state in which the variable \cf{sold} has value 1. This variable is set when \cf{factory} performs action \cf{sell} (see Line~\ref{lst:linesold} in Figure~\ref{fig:qflan_transition_system_VM}).  
We also set the two parameters specifying the required confidence interval, $\alpha$, and $\delta$, to 0.05 (for the price we use $\delta=0.5$). Therefore, we ask \mv{} to compute 95\% confidence intervals of width at most 0.05 (0.5 for price).

As a second set of experiments, we study the same properties, but at the varying of the simulation steps, from 1 to 500. This is obtained using Query 2 in Figure~\ref{fig:qflan_query}. We do this because this analysis regards a larger portion of the dynamics of the model, allowing us to further exemplify the advantages brought by our methodology.

\subsection{Results} \label{results: Effectiveness}
Considering Query 1, \mv{} instructed the QFLan probabilistic simulator to perform 1440 and 1600 simulations to estimate all properties for the case of maximum price 10 and 15, respectively.


\begin{table}[t]
\centering
\resizebox{0.7\textwidth}{!}{
    \begin{tabular}{cccccc}
    \toprule
    &\multicolumn{5}{c}{\emph{Studied properties}} 
    \\
    \cmidrule(l){2-6}
    \emph{Maximum Price} & \textbf{Avg Price} & \textbf{Tea}  & \textbf{Coffee} & \textbf{Cocoa} & \textbf{Cappuccino} \\
    \midrule 
    10 & 5.53 & 0.64 & 0.33 & 0.19 & 0.00 \\
    15 & 7.45 & 0.54 & 0.46 & 0.42 & 0.22 \\
    \bottomrule
\end{tabular}
}
\caption{\label{tab:QFLan_statistics_results} Numerical results experiments Vending machine experiments}
\end{table}

Table~\ref{tab:QFLan_statistics_results} lists the numerical results obtained by \mv{} for the two considered maximum prices. As shown in the first row in Table~\ref{tab:QFLan_statistics_results}, the results confirm that with maximum price 10 the probability of having \cf{Cappuccino} installed in sold machines is zero. 
Instead, for the case of a maximum price of 15, the probability increases to 0.2\rev{2}. This is in line with the results in~\cite{DBLP:conf/fm/VandinBLL18}. 
\rtwo{This is an example of potentially unexpected behavior resulting from the richness of QFLan's constraints: the model either allows or prohibits the feature \cf{Cappuccino} depending on the strictness of a constraint.}
Thanks to the methodology proposed in this paper, in addition to the numerical results, we can show graphically the behavior of the Vending machine model and how a different constraint on price can change the behavior of the model. 

\begin{figure}[t]
\centering
     \includegraphics[width=0.45\textwidth]{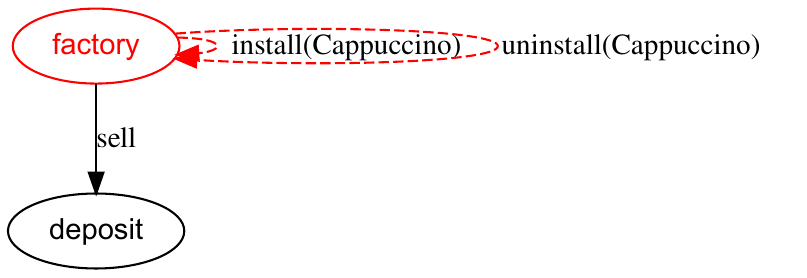}
     \hfill
      \includegraphics[width=0.45\textwidth]{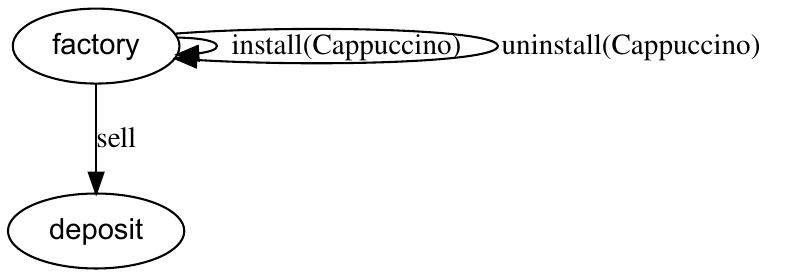}
\caption{Diff models for Query 1 of Figure~\ref{fig:qflan_query}. (Left) Model for maximum price 10. (Right) Model for maximum price 15. 
}
\label{fig:qflan_DFG_price_10_15_sold}
\end{figure}

Figure~\ref{fig:qflan_DFG_price_10_15_sold} depicts the diff models obtained by comparing the original QFLan model, and the ones mined using the simulation logs. 
Similarly to Figure~\ref{fig:qflan_model_behavior}, to improve readability we have edited the images to drop some edges irrelevant to this paper. We did not drop any red edges.
Figure~\ref{fig:qflan_DFG_price_10_15_sold} (Left)  refers to the case of maximum price 10. 
Here, we can see that the edges for \cf{install} and \cf{uninstall} of \cf{Cappuccino} are marked in red (as well as node \cf{factory}). This means that, even if those transitions are included in the model, they do not appear in its behavior (see Section~\ref{sec: method}). 
Figure~\ref{fig:qflan_DFG_price_10_15_sold} (Left) confirms our aforementioned hypothesis that the simulator never completes this transition because of the quantitative constraints. 

Figure~\ref{fig:qflan_DFG_price_10_15_sold} (Right) considers the case in which the maximum value for the price is 15. We can see that all the edges are now black. This means that the behavior of the model now also allows to install and uninstall \cf{Cappuccino}. We can, therefore, effectively execute all transitions present in the part of the model specification relevant to this query (i.e.\rev{,} the transitions from state \cf{factory}). 

\begin{figure}[t]
\hspace{-0.8cm}
\includegraphics[width=1.1\textwidth]{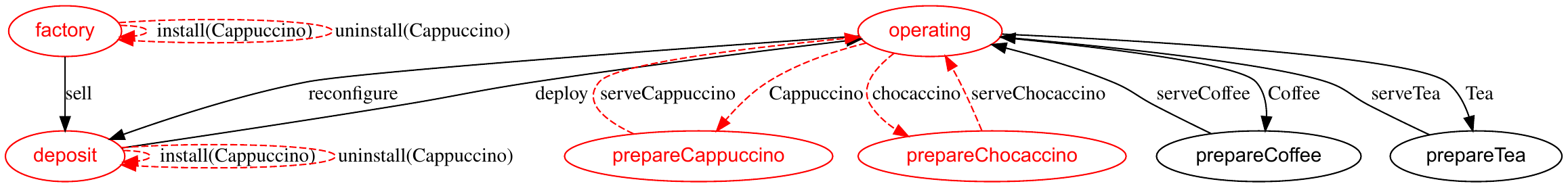}
\caption{Diff model for Query 2 of Figure~\ref{fig:qflan_query}  for maximum price 10. 
}
\label{fig:qflan_DFG_price_10_steps}
\end{figure}
\begin{figure}[t]
\hspace{-0.6cm}
\includegraphics[width=1.1\textwidth]{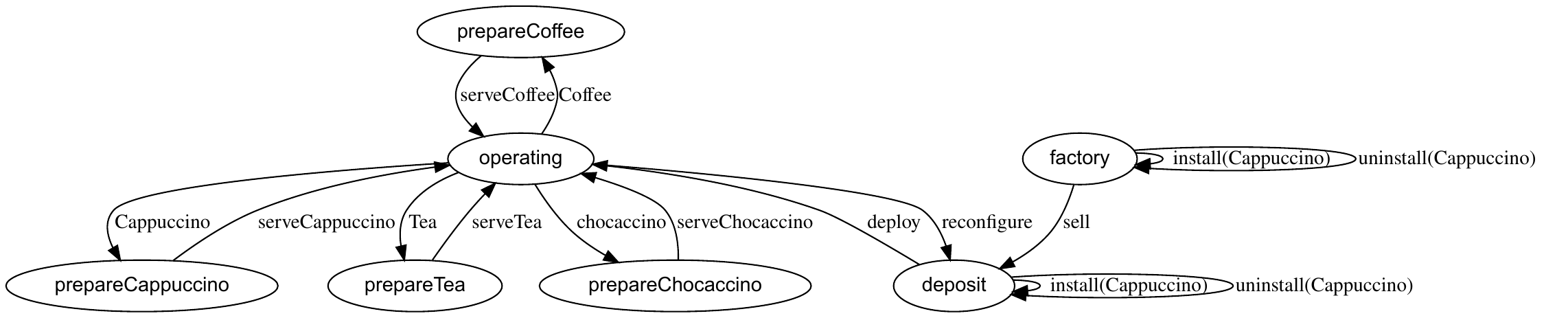}
\caption{Diff model for Query 2 of Figure~\ref{fig:qflan_query} for maximum price 15.}
\label{fig:qflan_DFG_price_15_steps}
\end{figure}

We now move our attention to Query 2 from Figure~\ref{fig:qflan_query}. This time we do not focus only on transitions executed in state \cf{factory}, but on all transitions executed in the first 500 simulation steps. We chose 500  because, from preliminary investigations, these are enough to allow the model to express all parts of its behavior. As for Query 1, we consider the two cases of maximum prices (10 and 15). 
For both experiments, \mv{} instructed the QFLan simulator to run 1440 simulations.


Figures~\ref{fig:qflan_DFG_price_10_steps} and~\ref{fig:qflan_DFG_price_15_steps} depict the diff models obtained by comparing the original model behavior, and the ones mined on the simulation logs. 
Figure~\ref{fig:qflan_DFG_price_10_steps} considers the case of maximum price 10, while Figure~\ref{fig:qflan_DFG_price_15_steps} 15. Images were edited similarly to Figure~\ref{fig:qflan_DFG_price_10_15_sold} to improve readability. 
We can see that the case of maximum price 10 presents several red edges (and nodes). Node \cf{factory} presents the missing transitions related to \cf{Cappuccino} discussed for Query 1. Furthermore, all transitions related to \cf{Cappuccino} in other nodes are missing as well. E.g., we never enter in state \cf{prepareCappuccino}, as we need to execute action \cf{Cappuccino} to get there, but this action is enabled only if the corresponding feature is installed. Likewise, we cannot serve \cf{Chocaccino}, because it requires to have both \cf{Cappuccino} and \cf{Cocoa} (see Figure~\ref{fig:qflan_Action_constraints}).

Conversely, Figure~\ref{fig:qflan_DFG_price_15_steps}  does not contain any red edge. Therefore, the more permissive constraint on maximum price does not prevent any part of the behavior of the machine. 

\paragraph{Discussion}
These experiments 
demonstrate the effectiveness of our method, which, thanks to the integration of SMC and PM techniques, \revsecond{enables us to thoroughly evaluate visually unexpected deviations between the behavior intended by the modeler, and the actual one obtained simulation the model. Such discrepancies might not be apparent in the model, as they might occur due to the richness of constraints present in QFLan. 
The application of PM techniques aids in exploring and uncovering these unexpected behaviors. }
  Therefore, we can positively answer to RQ1. 

\section{Experimental evaluation: RQ2 Scalability}
\label{subsec: rq2 scalability}
To answer RQ2, we consider a classic SPL example with parametric complexity: the elevator product line \rev{introduced in~\cite{PLATH200153}}. 
This has become a widely-used benchmark in PLE, especially as regards  scalability studies for novel methodologies \rev{(see, for instance,~\cite{dimovski2017efficient, DBLP:conf/icse/CordySHL13, DBLP:journals/fac/ChrszonDKB18, DBLP:journals/scp/ClassenCHLS14, DBLP:conf/icse/ApelRWGB13,DBLP:journals/tse/BeekLLV20,CHSL11,MBCCK16,MWKTS16,SJBK12}).}
\rev{This case study is considered particularly challenging by the community. It has $9$ independent and unconstrained features, yielding $512$~products. More importantly, it allows to consider instances with an increasingly large number of floors.}
In this section we use it with the same goal: to illustrate the scalability of our methodology.


\subsection{Domain description - Elevator}

\begin{figure}[t]
	\centering
    	 \includegraphics[width=0.8\textwidth]{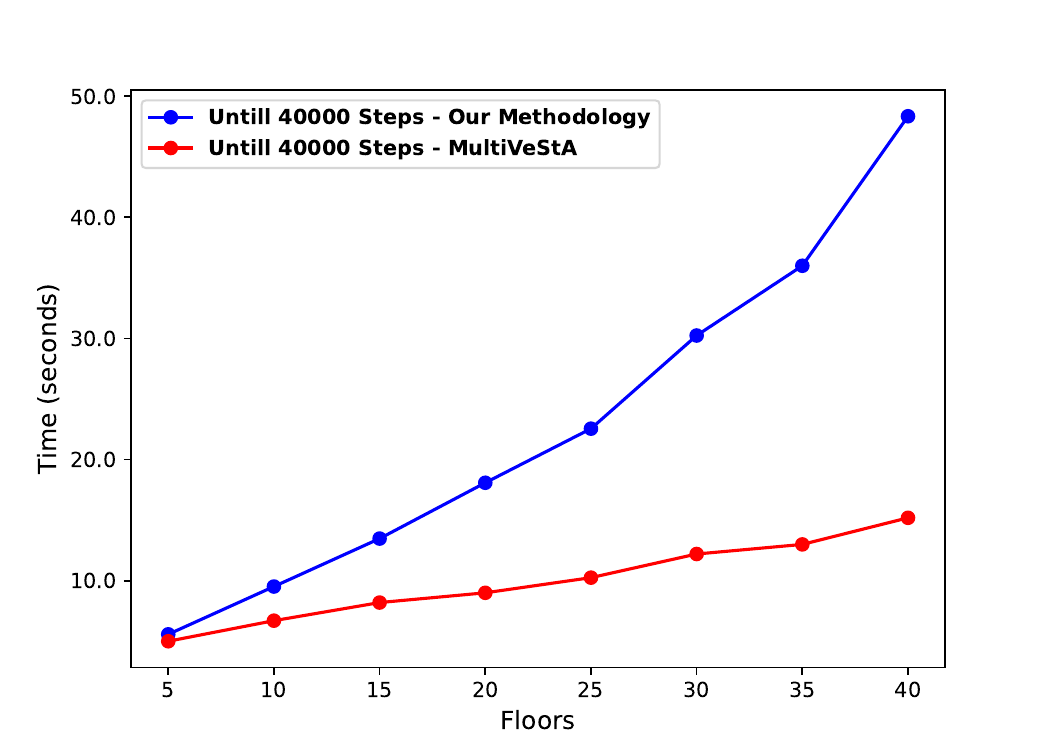}
\caption{Runtime analysis: Runtime of our methodology (blue line). Runtime of \mv{}  analysis (red line). Times are averaged over 10 replications of the analysis of the same \mv{} query over increasingly complex versions of the elevator SPL with 5, 10, \ldots, 40 floors.}
     \label{fig: Runtime analysis}
    \end{figure}

The elevator SPL has been provided in several \emph{incarnations}. Here we consider the one from~\cite{DBLP:journals/tse/BeekLLV20}. 
Product line engineers often encounter challenges when designing and developing products with numerous independent and unconstrained features. The Elevator SPL fits well for this scope. 
Furthermore, this case study is especially useful for product line engineers who need to consider instances of increasingly larger complexity, as it is possible to parameterize the model for different numbers of floors. 
The Elevator SPL consists of multiple platform and cabin buttons, one per floor, used to summon/direct the elevator. Once a button is pressed, it remains active until the elevator has served the corresponding floor by opening and closing its doors. The specific case study we consider focuses on nine key features that can alter the behavior of the elevator. For example, the feature \cf{AntiPrank} makes it necessary for a button to be kept pushed, while \cf{Park} makes the elevator return to the first floor when empty (see~\cite{DBLP:journals/tse/BeekLLV20} for the full list of considered features). 

In~\cite{DBLP:journals/tse/BeekLLV20}, the authors use this benchmark to study the scalability of QFLan at the varying of the number of floors from 5 to 40, while usual SPL approaches can scale up to 10 floors.
In particular, the authors of \cite{DBLP:journals/tse/BeekLLV20} consider the Elevator SPL with all unconstrained features from~\cite{PLATH200153}, with a fixed maximum capacity of the elevator set to eight persons, and a maximum allowed load of four persons. 

\subsection{Experiments}
%
We follow an approach similar to the one in~\cite{DBLP:journals/tse/BeekLLV20}. We consider a \mv{} property that checks that when the \cf{load} variable (representing the number of people in the elevator) exceeds the \cf{capacity} variable (representing the maximum capacity of the elevator), the elevator does not move (as indicated by variable \cf{direction} having value \cf{0.0}). To evaluate this property, \mv{} checks it for all states encountered within the first \cf{maxStep} steps. 
As soon as the condition is not satisfied, the current simulation terminates, otherwise, \cf{maxStep} steps are performed. 
In~\cite{DBLP:journals/tse/BeekLLV20}, the authors considered varying numbers of \cf{maxStep}, from $5,000$ to $40,000$. Here we consider only the latter largest case.
This property is always satisfied, i.e., we always get value 1. This was on purpose, to guarantee that every simulation will always consist of $40,000$ steps. 
%
As in~\cite{DBLP:journals/tse/BeekLLV20}, we consider a varying number of floors from 5 to 40. 

\subsection{Results}
The results are shown in Figure~\ref{fig: Runtime analysis}, providing in red the runtime of the \mv{} analysis, and in blue that of our methodology. The latter includes pre-processing of simulation logs, process mining, and generation of diff models.
We can see that our methodology succeeded in all instances and that it can provide results in less than a minute even for the largest instances. The methodology tends to have a runtime in the same order of magnitude of the \mv{} analysis, and it never exceeds more than 5 times its runtime. 

\paragraph{Discussion}
These experiments demonstrate the scalability of our method. In fact, it could be successfully applied to SPL models considered particularly challenging by the PLE community, and that are regularly used as benchmarks. Therefore, we can positively answer RQ2: our techniques can indeed be applied to large SPL models considered challenging by the PLE community. 


\section{Experimental evaluation: RQ3 Multi-domain}
\label{subsec: Multi-domain}
So far we have shown how our methodology can be applied to well-known SPL models from the PLE community.
The goal of this section is to show that our approach can easily generalize to models from other domains. In particular, we consider the cyber-security domain, using \emph{attack-defense tress} (ADT, or just attack trees). Their use is recommended by NATO~\cite{nato}, and are widely used, e.g., in aerospace~\cite{usdefense}, or safety-critical cyber-physical systems~\cite{HU2022107494}\rev{.}


\subsection{Domain description - Cyber-security Attacks and Threat Modeling}
We consider the threat model presented in~\cite{ter2021quantitative}, describing the attack strategies that a thief can attempt when trying to complete a robbery in a bank. 
\rtwo{Notably, the authors of~\cite{ter2021quantitative} emphasized that this model has infinite state space, preventing the use of exact analysis techniques based on an exhaustive exploration of such state space. The authors of~\cite{ter2021quantitative} leveraged this aspect to advocate the utilization of SMC for analyzing this model.}
In the preliminary workshop version of this paper~\cite{10.1007/978-3-031-25383-6_18}, we have exemplified how an embryonic version of our methodology could be applied to an extreme simplification of this model. 
%
The model has been created in RisQFLan~\cite{ter2021quantitative}, a member of the QFLan family recast to target the security domain. Similarly to QFLan, RisQFLan models consist of a declarative part, the \emph{attack-defense tree} (ADT), and a procedural part, the \emph{probabilistic attacker}. 
The intuition is that while a feature diagram describes a family of products, an ADT describes the family of possible attacks on a system. 
Similarly, while a probabilistic process of QFLan enables for dynamic reconfigurable SPLs, a probabilistic attacker in RisQFLan allows studying how vulnerable a system is to specific attackers (e.g.\rev{,} bound to stringent or permissive budget constraints). We refer to~\cite{ter2021quantitative} for a deep presentation of RisQFLan, and of how the feature-oriented framework QFLan has been recast to this new domain. 
%

\begin{figure}[t]
	\centering
    	 \includegraphics[width=0.6\textwidth]{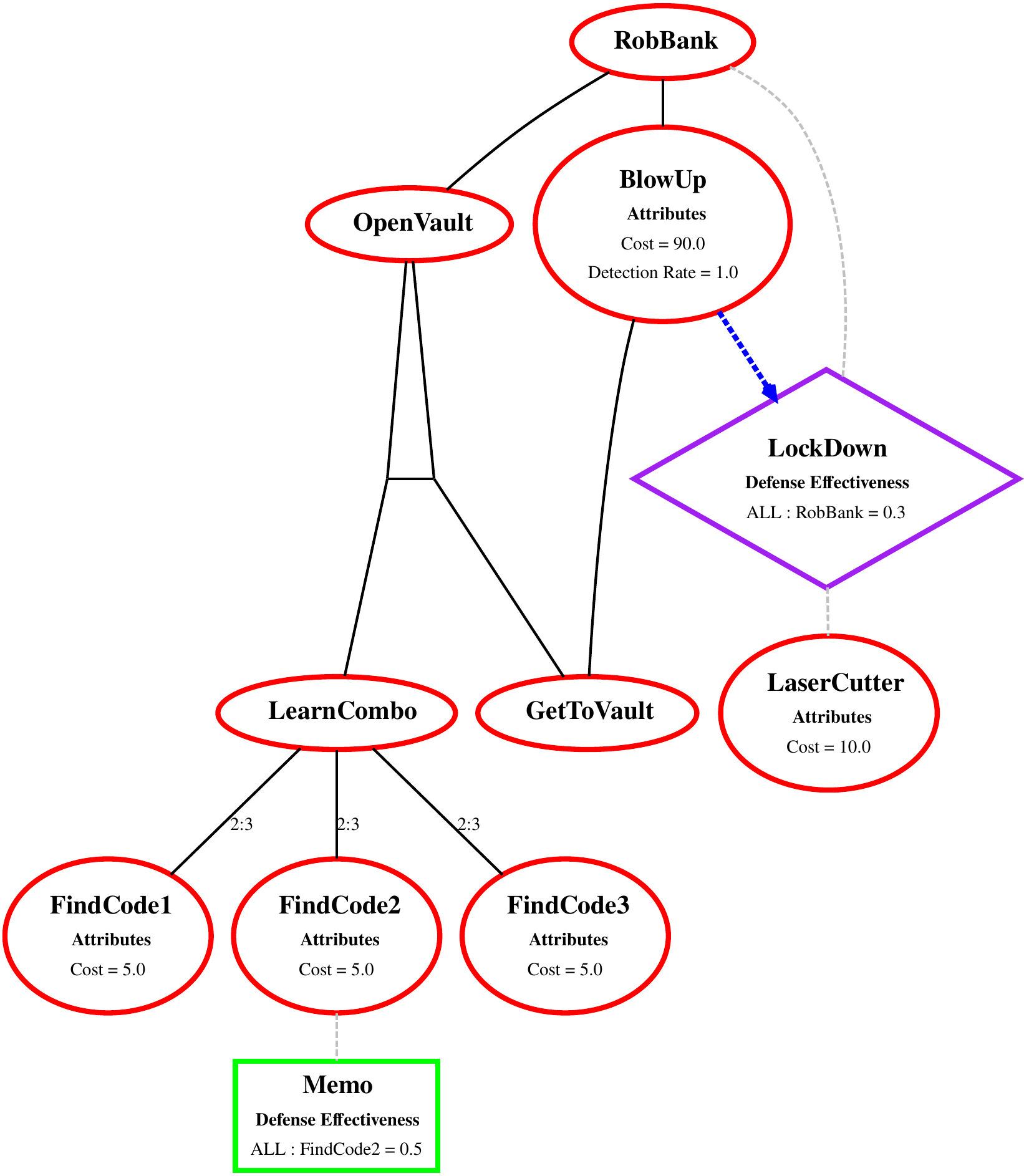}
\caption{ADT RobBank model}
     \label{fig:riqflan_ADT_robbabk}
    \end{figure}


Figure~\ref{fig:riqflan_ADT_robbabk} shows the ADT considered in this section. 
The root represents the root attack goal of robbing the bank. Nodes \cf{OpenVault} and \cf{BlowUp} are two sub-attacks, grouped by an \emph{OR}-relation. \cf{OpenVault} further refines in \cf{Learncombo} and \cf{GetToVault}, grouped by an \emph{AND}-relation: in order to open the vault, we need first to learn the combo and get to the vault.  \cf{LearnCombo} is further refined in three nodes, \cf{FindCodei}, grouped by a \emph{2-out-of-3}-relation. An attack node can succeed only if its refinements allow for it. E.g., \cf{LearnCombo} can succeed only if at least two \cf{FindCodei} attacks succeeded. 
An attack node in RisQFLan can be somehow mapped to a feature in QFLan, but these are different notions with different interpretations and therefore are handled differently in the formal semantics of the two languages. 
An ADT also has other types of nodes, \emph{defense} nodes, like the defense \cf{Memo} and the \emph{countermeasure} \cf{LockDown}. The former is a static defense that decreases the probability of success of the attacks to which it is connected (\cf{FindCode2}). Instead, a countermeasure is a dynamic defense that must be activated by attack attempts that it can monitor (\cf{BlowUp}), denoted by the blue arrow. Once activated, countermeasures behave like defenses. 

Intuitively, the ADT can be read as follow: the thief can attempt the robbery by two strategies: opening the vault or blowing it up. Both strategies require that the thief gets to the vault.  In addition,  opening the vault requires the thief to learn the combination of the vault, which in turn requires discovering at least two codes. 

A RisQFLan model is given in textual format, similar to QFLan. The complete model can be found in~\cite{ter2021quantitative}. Here we provide only the parts relevant to the performed experiments. 

%

A model in RisQFLan, similarly to QFLan, can be equipped with several
(quantitative) predicates and constraints. For example, Figure~\ref{fig:riqflan_ADT_robbabk} shows that attack nodes have a \cf{Cost}, paid by the attacker every time the corresponding sub-attack is attempted. At the same time, The block \cf{quantitative constraints} in Figure~\ref{fig:riqflan_transition} depicts how we can constrain the dynamics of an attacker to finite resources, e.g.\rev{,} by imposing attackers to not spend more than 100 (EUR).
In other words, the sum of the costs of attempted attacks cannot be higher than 100. 

\begin{figure}
	\centering
\begin{lstlisting}
    begin quantitative constraints
        { value(Cost) <= 100 }
    end quantitative constraints	
    begin actions
        tryAction tryGTV choose
    end actions
    
    begin attacker behaviour
    begin attack
        attacker = Thief
        states = Start, TryOpenVault, TryLearnCombo, TryFindCode, TryGetToVault, TryBlowUp, Complete 
        transitions = 
        Start - (succ(RobBank), 2, allowed(RobBank)) -> Complete,
        Start - (fail(RobBank), 1, allowed(RobBank)) -> Complete,
        //Get to the vault attempt
        Start -(tryGTV, 4, !has(GetToVault)) -> TryGetToVault,
        TryGetToVault -(succ(GetToVault) , 2, {AttackAttempts = AttackAttempts + 1}) -> Start,
        TryGetToVault -(fail(GetToVault), 1, {AttackAttempts = AttackAttempts + 1}) -> Start,
        //Open the vault attempt
        Start -(choose, 4) -> TryOpenVault,
        TryOpenVault -(succ(OpenVault) , 2, {AttackAttempts = AttackAttempts + 1},has(LearnCombo) and has(GetToVault)) -> Start,
        TryOpenVault -(fail(OpenVault), 1, {AttackAttempts = AttackAttempts + 1},has(LearnCombo) and has(GetToVault)) -> Start,
        TryOpenVault -(tryAction , 2, has(LearnCombo) and !has(GetToVault)) -> Start,
        TryOpenVault -(tryAction, 5, !has(LearnCombo)) -> TryLearnCombo,
        //Learn the combinations of he vault attempt
        ...
        //Blow up the vault attempt
        Start -(choose, 4) -> TryBlowUp,
        ...
    end attack
    end attacker behaviour    
    
    begin init
        // LockDown cannot be activated if we have LaserCutter
        Thief = {FindCode1, LaserCutter }
    end init
\end{lstlisting}
\caption{Probabilistic attacker in RisQFLan}
     \label{fig:riqflan_transition}
    \end{figure}
The considered probabilistic attacker is illustrated in Figure~\ref{fig:riqflan_transition}, given in a format similar to that of the probabilistic process of QFLan. 
Similarly to QFLan, the RisQFLan simulator will select the next transition to execute in a simulation depending on their weights and on several types of constraints. 

\begin{figure}[t]
	\centering
    	 \includegraphics[width=1\textwidth]{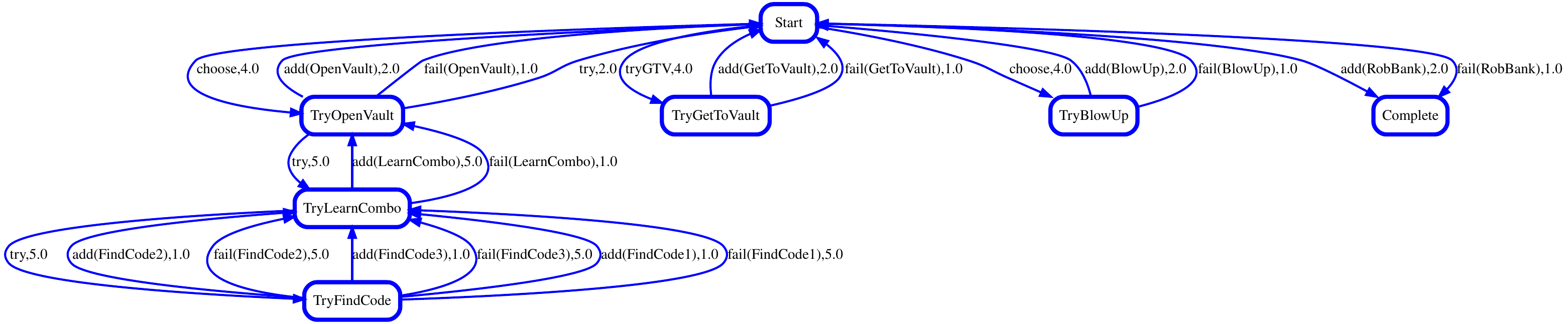}
\caption{RisQFLan model: Attacker behavior}
     \label{fig:riqflan_robbank_attacker_bahavior_original}
    \end{figure}

In the model, we can also define the prior knowledge and availability of the attacker. The block \cf{init} in Figure~\ref{fig:riqflan_transition} imposes that the thief already knows the first combination, and owns a laser cutter to disable the lockdown defense. 

Similarly to Figure~\ref{fig:qflan_model_behavior}, Figure~\ref{fig:riqflan_robbank_attacker_bahavior_original} provides a graphical representation of the probabilistic attacker.
This plays the role of the input model in Figure~\ref{fig:abstract_example}. 
From Figure~\ref{fig:riqflan_robbank_attacker_bahavior_original} we can see that each attack begins in the \cf{Start} node, where the simulator can choose to blow up or open the vault. After an attack attempt, the simulator returns to the Start state, where it chooses to try another attack or to complete the robbery if this is allowed. If the root attack succeeds, the attacker will move, and terminate, in state \cf{Complete}. 
Attack attempts might also fail. This is dictated by the weights added in the probabilistic attacker and by the existing defenses. 

\subsection{Experiments on the original model}
\label{subsec: Experiments_original_model_risqflan}

\begin{figure}[t]
	\centering
\begin{lstlisting}
 begin analysis
		query = eval from 1 to 100 by 1 :
		{RobBank, OpenVault, BlowUp,LearnCombo, GetToVault, FindCode2, FindCode3,LockDown}
		default delta = 0.1		alpha = 0.1		parallelism = 1
		logs = "log_RobBank.csv"
 end analysis
\end{lstlisting}      
\caption{Query to invoke \mv\ RobBank model}
     \label{fig:riqflan_query}
    \end{figure}

To demonstrate that our method can automatically discover unwanted behaviors also in this domain, we use \mv{} to analyze the query in 
Figure~\ref{fig:riqflan_query}. This instructs \mv{} to evaluate the probability of success of eight attacks in each simulation step from 1 to 100. The CI specification is as in QFLan. 
As in~\cite{ter2021quantitative}, we assume that the attacker owns a \cf{LaserCutter} which disables the \cf{Lockdown} defense, and that s/he already succeeded in obtaining the first code of the vault (Figure~\ref{fig:riqflan_transition}). We will show that our methodology can pinpoint issues in the model, and how it can suggest fixes. 

\subsection{Results on the original model}
\label{subsec: results_original_model_risqflan}
\mv{} instructed the probabilistic simulator to run 320 simulations. 
Table~\ref{tab:risqflan_statistics_results_robbank} lists the analysis results obtained for step 100. We can notice that the probability of a total lockdown is equal to zero, as expected by the presence of the \cf{LaserCutter}. 
\begin{table}[t]
\resizebox{14cm}{!}{
\begin{tabular}{c c c c c c c c c}
\toprule
\textbf{Property} & RobBank & OpenVault & BlowUp & LearnCombo & GetToVault & FindCode2 & FindCode3 & LockDown \\ \midrule
\textbf{Probability} & 0.175   & 0.204     & 0.062  & 0.515      & 0.525      & 0.221     & 0.363     & 0.0 \\ \bottomrule
\end{tabular} }
\caption{\label{tab:risqflan_statistics_results_robbank} Numerical results experiments on the original RobBank model}
\end{table}

Another critical element highlighted by the SMC analysis is that the probability of succeeding in the root attack is 0.175 despite \cf{LockDown} is disabled.\rev{ } 
The reasons why only a few simulations (about 18\%) ended with a complete robbery of the bank are not easy to spot just by a black-box inspection of the numerical results. 
\begin{figure}[t]
    \centering
    \includegraphics[width=1\textwidth]{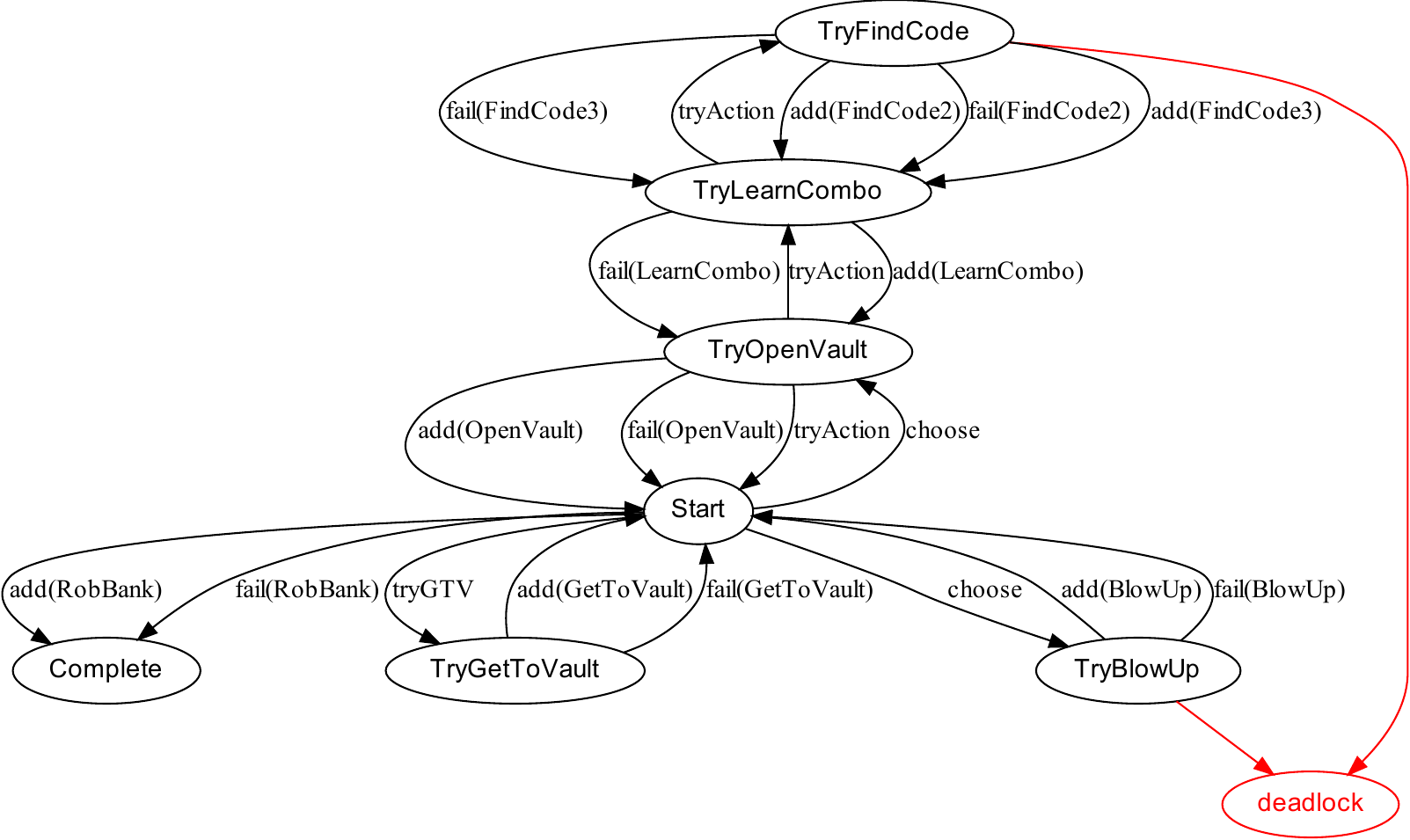}
    \caption{Diff model obtained experiments on the original RobBank model}
    \label{fig:riqflan_DFG_robbank_original}
\end{figure}

Figure~\ref{fig:riqflan_DFG_robbank_original} depicts the diff model produced by our methodology. 
It 
displays two red edges and a red node; all the remaining edges and nodes are colored in black. 
In particular, the red node is a special node added by our methodology: a \emph{deadlock} node. It denotes 
simulations that ended unexpectedly because no transitions were enabled. 
Thanks the two red edges, we can see that some of the simulations ended unexpectedly in the states \cf{TryFindCode} and \cf{TryBlowUp}, lowering the overall success probability. 

\paragraph{Issue in the model, and fix}
Figure~\ref{fig:riqflan_DFG_robbank_original} highlights issues in states \cf{TryFindCode} and \cf{TryBlowUp}. By looking at the original model specification, we can see that these states can only perform transitions to attempt attacks (transitions \cf{fail} and \cf{add} in Figure~\ref{fig:riqflan_robbank_attacker_bahavior_original}). The execution of these transitions increases a cost given by the cost of the attempted attacks (5 for \cf{Findcode}, 90 for \cf{BlowUp}). This makes us suspect that the deadlocks are due to the constraint on the maximum cost being set to only 100 (Figure~\ref{fig:riqflan_transition}). 
Therefore, if the attacker gets into these states without enough money to attempt the corresponding attack, s/he will get stuck there because no transition is enabled. 
To fix this unwanted behavior and to have a more reliable evaluation of the properties, the modeler needs to refine it by adding an extra \emph{escape} transition in each of these two nodes to go back to their direct parent nodes. This transition shall have no cost, which is obtained by executing a custom action (see Figure~\ref{fig:riqflan_transition}). 

\subsection{Refined model}
\label{subsec: refined_model_risqflan}
\begin{figure}[t]
    \centering
    \includegraphics[width=1\textwidth]{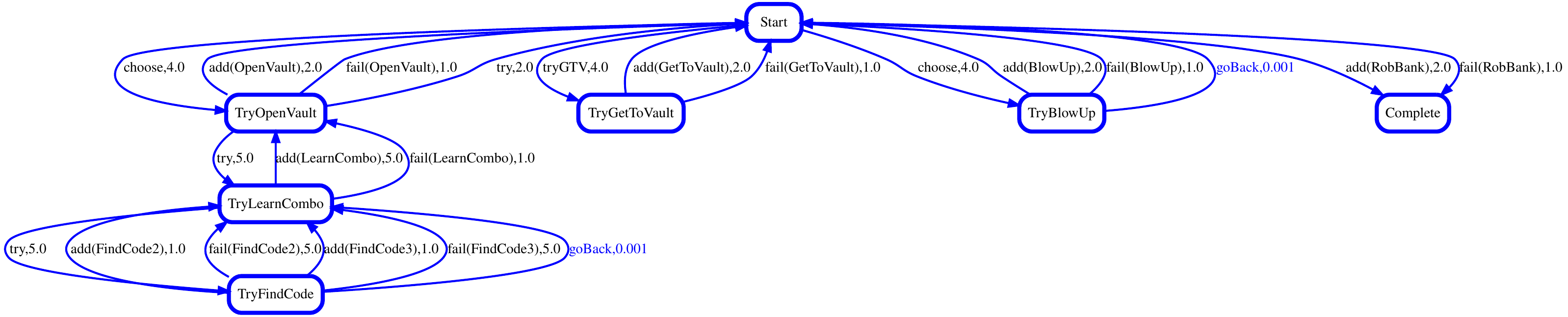}
    \caption{RisQFLan model: probabilistic attacker - Refined model}
    \label{fig:riqflan_robbank_attacker_bahavior_refined}
\end{figure}

We refine our model by adding a new action \cf{goBack} in Figure~\ref{fig:riqflan_transition}, and two transitions with this action from \cf{TryBlowUp} to  \cf{Start},  and from \cf{TryFindCode} to \cf{TryLearnCombo}.  Figure~\ref{fig:riqflan_robbank_attacker_bahavior_refined} depicts the refined attacker behavior with the new transitions highlighted in blue. We assign a low weight to these transitions, i.e., 0.1, to ensure that the simulator tends to choose them only when other options are not permitted.\footnote{\rtwo{The use of a low probability is a  workaround. We could have used so-called \emph{action constraints}~\cite{ter2021quantitative}, but this would have required an in depth description of RisQFLan.}}  

\begin{figure}[t]
    \centering
    \includegraphics[width=1\textwidth]{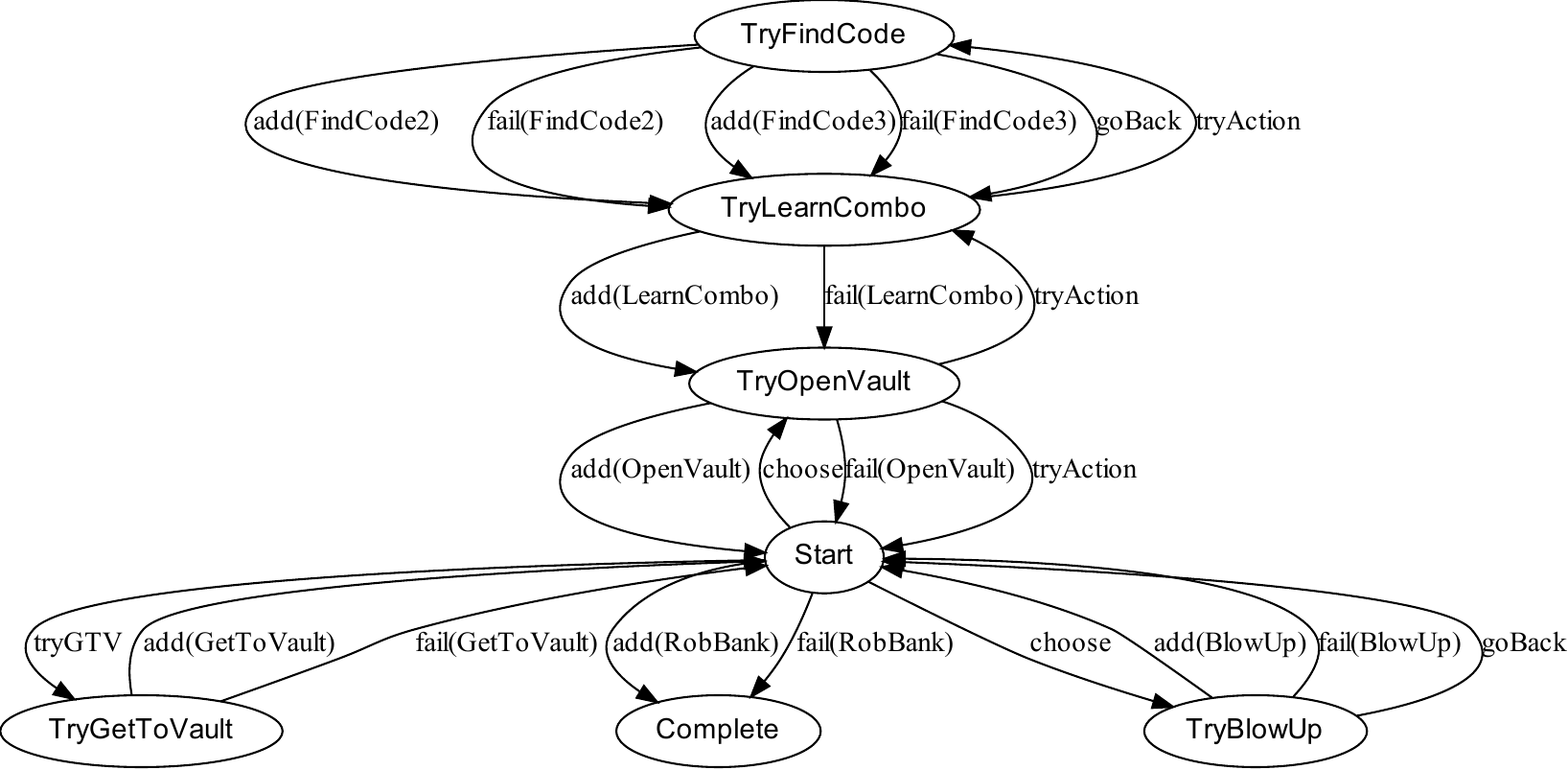}
    \caption{Diff model for the refined RobBank model}
    \label{fig:riqflan_DFG_robbank_refined}
\end{figure}

\paragraph{Experiments and results on the refined model}
\label{subsec: Experiments_refined_model_risqflan}
Thanks to these new transitions, the observed dynamics shall not exhibit anymore the discussed deadlocks. To ensure this, we use the same query from Figure~\ref{fig:riqflan_query}. Also, in this case, \mv{} required to run 320 simulations.
The results are given in Table~\ref{tab:risqflan_statistics_results_robbank_refined_model}. Besides \cf{LockDown}, which is again equal to zero, all the other properties increased. 
The obtained diff model is given in 
Figure~\ref{fig:riqflan_DFG_robbank_refined}. No red edges or nodes are present, 
meaning that we fixed the issues. 


\begin{table}[t]
\resizebox{14cm}{!}{
\begin{tabular}{c c c c c c c c c}
\toprule
\textbf{Property} & RobBank & OpenVault & BlowUp & LearnCombo & GetToVault & FindCode2 & FindCode3 & LockDown \\ \midrule
\textbf{Probability} & 0.393   & 0.572     & 0.12  & 0.59      & 0.825      & 0.21     & 0.453     & 0.0 \\ \bottomrule
\end{tabular} }
\caption{\label{tab:risqflan_statistics_results_robbank_refined_model} Numerical results experiments on the refined RobBank model}
\end{table}

\paragraph{Discussion}
These experiments demonstrate that our methodology can be applied to domains beyond SPL, including cyber-security ones. \revsecond{Therefore, we can answer positively to RQ3}, since we have automatically discovered unwanted and unexpected behaviors. In addition, we also got hints on how to fix such issues by obtaining a refined model that does not show the issues. 

%



%% file: relatedwork.tex
\section{Related work}
\label{sec:related}

Since our approach is centered around the utilization of automated verification techniques, specifically probabilistic model checking combined with PM techniques, within the specific context of behavioral models of dynamic SPLs, we will give an overview of the related works that used models for specifying SPLs, providing the behavior of those along with the related verification techniques and tools. 
The well-known behavioral modeling languages for SPLs rely on overlaying multiple labeled transition systems (LTSs) that represent different variants of products onto a single, augmented LTS family model. However, \revsecond{only a few~\cite{DBLP:journals/fac/ChrszonDKB18,DBLP:journals/taosd/DubslaffBK14}} of these languages enable the specification of probabilistic SPL models, and \rtwo{even less} support model-checking approaches. 
Featured Transition Systems (FTSs) were first introduced in~\cite{CHSLR10} and later expanded upon in~\cite{DBLP:conf/icse/ClassenHSL11} and~\cite{DBLP:series/lncs/CordyCHLS13}. An FTS represents a family of LTSs, with each LTS corresponding to a specific product. The LTSs are derived by projecting feature expressions (Boolean formulas defined over the feature set) assigned to the transitions. Transitions whose feature expressions are not satisfied by a particular product's feature set are eliminated, along with any unreachable states and transitions. In QFLan~\cite{DBLP:journals/tse/BeekLLV20}, the action constraints are similar to feature expressions in FTSs but they are applied to actions rather than transitions. While feature expressions offer more fine-grained specifications, action constraints in QFLan provide a more concise and declarative approach and support more general constraints and accommodate the modeling of adaptive or dynamic SPLs compared to FTS feature expressions. Modal Transition Systems (MTSs)~\cite{DBLP:conf/issta/FischbeinUB06, DBLP:journals/jlp/BeekFGM16} represent a family of LTSs that, similar to FTSs, distinguishes between admissible transitions and necessary transitions. Nevertheless, in comparison with this family of LTSs, QFLan provides support for feature attributes and richer quantitative constraints and allows for the modeling of dynamic SPLs since the feature set is statically determined upfront. In~\cite{DBLP:journals/ese/BeekDLMP22}, unreachable states and transitions, so-called hidden deadlocks, are made explicit through an algorithm that effectively transforms ambiguous FTSs into unambiguous ones and transforms them into a Modular Transition System that the modeler can more efficiently check. With QFLan, we are not interested in the static analysis of the formal model as in~\cite{DBLP:journals/ese/BeekDLMP22}, but, instead, we run the model and conduct a sufficient number of simulations to attain statistically significant outcomes regarding the particular query under investigation. 

A sequence of works, summarized in~\cite{DBLP:conf/isola/LeuckerT12}, introduced the notion of Product Line CCS (PL-CCS). PL-CCS expands upon the CCS framework by incorporating a variant operator, which enables the representation of alternative behaviors as alternative processes. The objective is to ensure the existence of only one of these processes during runtime. Another notable approach, described in~\cite{DBLP:journals/tosem/ErwigW11}, is the choice calculus, which creates a common language for software variation management and aims to establish a foundational model for software variation, similar to the lambda calculus in programming languages. Additionally, DeltaCCS~\cite{DBLP:journals/jlp/LochauMBR16}, an extension of CCS, draws inspiration from the widely used delta-modeling approach employed in automated product derivation for SPLs. The approach discussed in~\cite{DBLP:conf/gpce/ClarkeHS10} utilizes deltas to specify incremental changes to a core product. In contrast to PL-CCS and the choice calculus, DeltaCCS follows a modular approach in which choices are applied at well-defined variation points. Model-checking algorithms have been implemented in MAUDE to verify SPLs specified in DeltaCCS against modal m-calculus formulas. Despite both PL-CCS and DeltaCCS offering fundamental mechanisms for restructuring or modifying SPLs, they are not able to effectively model dynamic SPLs.  An alternative technique, known as Variant Process Algebra (VPA), is proposed in~\cite{DBLP:conf/splc/Tribastone14} for formal reasoning about SPLs but places emphasis on behavioral (bi)simulation relations rather than verification through model checking. For our work, we use FLan, a feature-oriented language designed to demonstrate how to specify both declarative and procedural aspects of product families. FLan draws inspiration from concurrent constraint programming, combining a store of constraints for declaratively expressing common constraints on features, including cross-tree constraints found in feature models. Additionally, FLan offers a comprehensive set of process-algebraic operators to procedurally specify product configurations and behaviors by supporting a wide range of constraints that can encompass quantitative aspects of feature attributes.
Family-based model checking of behavioral SPL models offers a powerful approach to simultaneously verify multiple behavioral product models within a single run. This technique enables the verification of properties using specialized SPL model-checking tools such as SNIP~\cite{DBLP:journals/sttt/ClassenCHLS12}, ProVeLines~\cite{DBLP:conf/splc/CordyCHSL13}, VMC~\cite{DBLP:conf/fm/BeekMS12, DBLP:journals/jlp/BeekFGM16, DBLP:conf/splc/BeekM14}, fNuSMV~\cite{DBLP:journals/scp/ClassenCHLS14, DBLP:conf/fase/DimovskiLW19}, and ProFeat~\cite{DBLP:journals/fac/ChrszonDKB18} for probabilistic model checking. \revsecond{ProFeat is an example of  technique that utilizes numerical computations to achieve precise outcomes when evaluating the properties of a model.} 
Alternatively, traditional model checkers such as SPIN~\cite{10.5555/2951604, DBLP:journals/sttt/DimovskiABW17, DBLP:conf/fase/DimovskiW17}, PRISM~\cite{DBLP:journals/taosd/DubslaffBK14} for probabilistic model checking, Maude~\cite{DBLP:journals/jlp/LochauMBR16}, mCRL2~\cite{DBLP:conf/fase/BeekVW17, DBLP:journals/tse/BeekLLV20}, or NuSMV~\cite{DBLP:journals/sttt/Dimovski20} can be employed through appropriate abstractions or encodings. These classical model checkers can effectively verify properties of SPL models by leveraging suitable transformations or encodings to adapt them for SPL-specific analyses. In comparison to conventional product-based model checking approaches, QFLan's statistical model-checking features provide several noteworthy benefits. Firstly, the process of performing simulations can be effortlessly parallelized and distributed across multiple cores, clusters, or distributed computing systems, resulting in nearly linear improvements in processing speed. This parallelization capability enables significant acceleration of the overall verification process. Secondly, the same set of simulations can be utilized to evaluate multiple properties simultaneously, leading to a reduction in the computational time required for verifying each property individually. This simultaneous checking of multiple properties further enhances the efficiency of the verification process. 
About enhancing SMC techniques with PM techniques, besides our preliminary work~\cite{10.1007/978-3-031-25383-6_18}, directed to demonstrate the potentiality of these techniques applied on a threat model, to best of our knowledge there are no other previous works that apply PM techniques to probabilistic model checking on SPLs models. 

%% file: discussion.tex
\section{Conclusion}
\label{sec:disc}
We presented a novel approach for the validation of simulation models, and in particular software product lines from product lines engineering.
The methodology consists of a combination of simulation-based analysis techniques from statistical model checking (SMC)~\cite{Agha18}, and process-oriented data-driven techniques from process mining (PM)~\cite{VanderAalst2016}.
In particular, we use PM to explain SMC analyses, obtaining a graphical representation of the system behavior as observed in the SMC simulations. In our experimental evaluation, we demonstrate that: (1) our methodology helps in identifying issues in the model, and in getting hints on how to fix them (2) it scales to complex models, and (3) it is general because it can be applied to domains beyond product line engineering.


In future investigations, we will investigate if PM can further help in the product line engineering. For example, it might be useful to compare different products from the same family. %
Furthermore, we will also study whether PM can further help in solving issues of SMC, e.g., how to handle rare events~\cite{DBLP:conf/rp/LegayST16}. 
%
We already considered two different domains, namely product lines engineering, and risk modeling and analysis, analyzed for SMC. In the future, it might be interesting to consider further types of probabilistic models and of analysis techniques like, e.g., pGCL programs studied with used in probabilistic model checkers like STORM~\footnote{\url{https://www.stormchecker.org/documentation/background/languages.html\#cpgcl}}. 
Finally, we plan to investigate the application of our methodology to further frameworks for dynamic PL models. A notable example is ProFeat~\cite{DBLP:journals/fac/ChrszonDKB18}. It is built on top of the probabilistic model checker PRISM~\cite{DBLP:conf/cav/KwiatkowskaNP11}, and  allows as well for SMC-based analyses. 
%


%% file: appendix.tex
\appendix

\section{Addressing typos in models}\label{sec:appendix}
\begin{figure}
    \centering
    \includegraphics[width=1\textwidth]{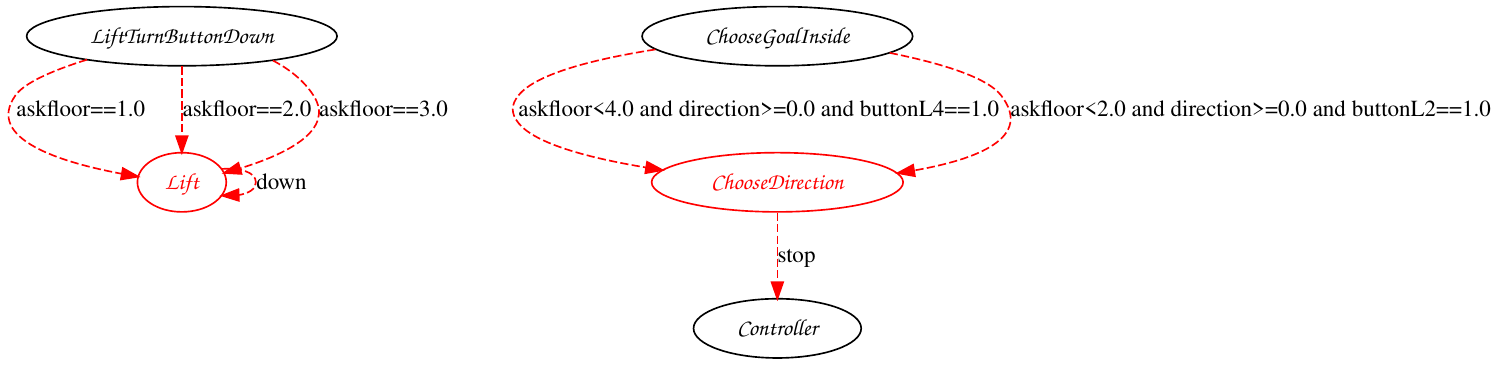}
    \caption{Excerpt of the diff model for the elevator model with 5 floors with the typo.}
    \label{fig: diff_elevator_reduced_5}
\end{figure}

\revsecond{Figure~\ref{fig:qflam_transition_elev_proccesses} lists a partial probabilistic process of the elevator model with 5 floors and involving four concurrent processes. This example highlights how a minor typo that could easily be overlooked when specifying the model, especially in complex models like the elevator, might be spotted by our methdology. 
The typo is in line~\ref{lst:weight_typo}, where, instead of adding a weight greater than zero, it is erroneously set to zero. As expected, the natural consequence of this typo is that the simulator will not traverse that transition, resulting in unexpected behavior and therefore biased  SMC analysis. Figure~\ref{fig: diff_elevator_reduced_5} illustrates the diff model, displaying only the relevant sections of the four concurrent processes. Among these sections, there exists at least one transition that the simulator did not traverse due to the typo in line~\ref{lst:weight_typo}, identifiable by the dashed red edge.}
\revsecond{The original model and the full diff model generated by our methodology can be accessed in \url{https://doi.org/10.5281/zenodo.8362717}. 
}

\begin{figure}
	\centering
\begin{lstlisting}
begin processes diagram
	begin process LiftProc //... Lift Process
	states = Lift,LiftTurnButtonDown
	transitions=
    	Lift -(open,1 ,{door=1})-> LiftTurnButtonDown,
    	Lift -(close,1 ,{door=0})-> Lift,
    	Lift -(up,1 ,{floor=floor + 1})-> Lift,
    	Lift -(down,0 ,{floor=floor - 1})-> Lift,(*\label{lst:weight_typo}*)
    	Lift -(clean,100 ,{buttonL0=0, 
                buttonL1=0,buttonL2=0, 
                buttonL3=0, buttonL4=0})-> Lift,

	LiftTurnButtonDown -(ask({floor==0}),100,{buttonL0=0,buttonF0= 0})-> Lift,
	LiftTurnButtonDown -(ask({floor==1}),100,{buttonL1=0,buttonF1= 0})-> Lift,
	LiftTurnButtonDown -(ask({floor==2}),100,{buttonL2=0,buttonF2= 0})-> Lift,
	LiftTurnButtonDown -(ask({floor==3}),100,{buttonL3=0,buttonF3= 0})-> Lift,
	LiftTurnButtonDown -(ask({floor==4}),100,{buttonL4=0,buttonF4= 0})-> Lift
	end process
 
 begin process ControllerProc
    //... Controller Process
    end process
    
 begin process ButtonsProc
    //... Button Process
     end process
 
 begin process PeopleProc
    //... People Process
    end process

\end{lstlisting}
\caption{Probabilistic processes of the Elavator model in QFLan}
     \label{fig:qflam_transition_elev_proccesses}
    \end{figure}

